\documentstyle[12pt,epsfig,supertab,rotating]{article}                                                                  
\def\3{\ss}                                                                                        

\begin{document}                                                                 
\thispagestyle{empty}
\title{\bf Measurement of the Proton Structure Function {\boldmath $F_2$} and
{\boldmath $\sigma_{tot}^{\gamma^* p}$}
at Low {\boldmath $Q^2$} and Very Low {\boldmath $x$} at HERA \vspace{0.5in}}
\author{ ZEUS Collaboration }
\date{}
\maketitle
\vspace{5 cm}
\begin{abstract}
A small electromagnetic sampling 
calorimeter, installed in the ZEUS experiment in 1995, significantly 
enhanced the acceptance for very low $x$ and low $Q^2$ 
inelastic neutral current scattering, $e^{+}p\rightarrow  e^{+}X$, at HERA. 
A measurement of the proton structure function $F_2$ and the total 
virtual photon-proton ($\gamma^*p$) cross-section is presented for 
$0.11 \le Q^{2} \le 0.65$  GeV$^2$ and 
$2 \times 10^{-6} \le x \le  6 \times 10^{-5}$, 
corresponding to a range in the  $\gamma^{*}p$ c.m. energy of
$100 \le W \le 230$ GeV. Comparisons with 
various models are also presented.

\end{abstract}                                                   %
\vspace{-19.5cm}
\begin{flushleft}
\tt DESY 97-135\\
July 1997 
\end{flushleft}
\thispagestyle{empty}

\newpage
\pagestyle{plain}
\setcounter{page}{1}
\begin{center}                                                                                     
{                      \Large  The ZEUS Collaboration              }                               
\end{center}                                                                                       
  J.~Breitweg,                                                                                     
  M.~Derrick,                                                                                      
  D.~Krakauer,                                                                                     
  S.~Magill,                                                                                       
  D.~Mikunas,                                                                                      
  B.~Musgrave,                                                                                     
  J.~Repond,                                                                                       
  R.~Stanek,                                                                                       
  R.L.~Talaga,                                                                                     
  R.~Yoshida,                                                                                      
  H.~Zhang  \\                                                                                     
 {\it Argonne National Laboratory, Argonne, IL, USA}~$^{p}$
% ...
\newpage                                                                                           
% ...
$^{\    2}$ now at Univ. of Salerno and INFN Napoli, Italy \\                                      
% ...
$^{  23}$ now at ATLAS Collaboration, Univ. of Munich \\                                           
% ...
$^{  32}$ partially supported by the Foundation for German-Russian Collaboration DFG-RFBR (grant nos 436 \\
\hspace*{3.5mm}RUS 113/248/3 and 436 RUS 113/248/2) \\
% ...
                                                           %
                                                           %
\newpage   
                                                           %
                                                           %
\begin{tabular}[h]{rp{14cm}}                                                                       
$^{a}$ &  supported by the Natural Sciences and Engineering Research                               
          Council of Canada (NSERC)  \\                                                            
$^{b}$ &  supported by the FCAR of Qu\'ebec, Canada  \\                                            
$^{c}$ &  supported by the German Federal Ministry for Education and                               
          Science, Research and Technology (BMBF), under contract                                  
          numbers 057BN19P, 057FR19P, 057HH19P, 057HH29P, 057SI75I \\                              
$^{d}$ &  supported by the MINERVA Gesellschaft f\"ur Forschung GmbH,                              
          the German Israeli Foundation, and the U.S.-Israel Binational                            
          Science Foundation \\                                                                    
$^{e}$ &  supported by the German Israeli Foundation, and                                          
          by the Israel Science Foundation                                                         
  \\                                                                                               
$^{f}$ &  supported by the Italian National Institute for Nuclear Physics                          
          (INFN) \\                                                                                
$^{g}$ &  supported by the Japanese Ministry of Education, Science and                             
          Culture (the Monbusho) and its grants for Scientific Research \\                         
$^{h}$ &  supported by the Korean Ministry of Education and Korea Science                          
          and Engineering Foundation  \\                                                           
$^{i}$ &  supported by the Netherlands Foundation for Research on                                  
          Matter (FOM) \\                                                                          
$^{j}$ &  supported by the Polish State Committee for Scientific                                   
          Research, grant No.~115/E-343/SPUB/P03/002/97, 2P03B10512,                               
          2P03B10612, 2P03B14212, 2P03B10412 \\                                                    
$^{k}$ &  supported by the Polish State Committee for Scientific                                   
          Research (grant No. 2P03B08308) and Foundation for                                       
          Polish-German Collaboration  \\                                                          
$^{l}$ &  partially supported by the German Federal Ministry for                                   
          Education and Science, Research and Technology (BMBF)  \\                                
$^{m}$ &  supported by the Fund for Fundamental Research of Russian                                
          Ministry for Science and Edu\-cation and by the German Federal                           
          Ministry for Education and Science, Research and                                         
          Technology (BMBF) \\                                                                     
$^{n}$ &  supported by the Spanish Ministry of Education                                           
          and Science through funds provided by CICYT \\                                           
$^{o}$ &  supported by the Particle Physics and                                                    
          Astronomy Research Council \\                                                            
$^{p}$ &  supported by the US Department of Energy \\                                              
$^{q}$ &  supported by the US National Science Foundation \\                                       
\end{tabular}                                                                                      
                                                           %

\pagebreak
\pagenumbering{arabic}
\pagestyle{plain}
\setcounter{page}{1}

\section{Introduction}
An early discovery at HERA was the rapid rise of 
the proton structure function, $F_2(x,Q^2)$, as the Bjorken 
scaling variable, $x$, decreases at low $x$ for photon virtualities
$Q^2> 10$  GeV$^2$ \cite{ref-ZEUSfirst, ref-H1first}. The 
ZEUS \cite{f2_zeus} and H1 \cite{f2_h1} Collaborations have 
extended the measurement of $F_2$ down to a $Q^2$ value of 1.5 
GeV$^2$. One of the most interesting features of the recent data is 
the persistence to the lowest $Q^2$ of the rapid rise of 
$F_2$ with decreasing $x$. 
The predictions of Gl\"{u}ck, Reya and Vogt (GRV) \cite{GRV}, 
which result from the dynamic generation of parton densities 
via next-to-leading order (NLO) perturbative QCD 
(pQCD) DGLAP\cite{ref-DGLAP} evolution of valence type 
distributions starting at a very low scale, $Q^2_0 = 0.34$ 
GeV$^2$, are in broad agreement with this observation.  It is 
surprising that leading twist NLO pQCD can describe the data down
to $Q^2=1.5$ GeV$^2$. Other global fit analyses based on NLO 
pQCD, such as those performed by MRS \cite{mrsa} and CTEQ 
\cite{CTEQ}, typically have much higher starting scales 
 $Q^2_0 = 3$ -- 5 GeV$^2$. It then becomes an interesting 
question to determine at which $Q^2$ the behaviour of $F_2$ 
becomes dominated by non-perturbative contributions. 

$F_2$ is related to the total virtual photon-proton ($\gamma^*p$) 
cross-section by 
$\sigma_{tot}^{\gamma^{*}p} \approx  (4 \pi^2 \alpha /Q^2) F_2$ 
for $x \ll 1$.  At fixed $Q^2$, 
the rapid rise of $F_2$ with decreasing $x$ is equivalent to a rapid rise 
of the total $\gamma^*p$ cross-section with c.m. energy, 
$W$ ($W^2 \approx Q^2/x$ in this 
kinematic regime). At high $W$, $\sigma_{tot}^{\gamma^{*}p}$ 
can be described by a power law behaviour, 
$\sigma_{tot}^{\gamma^{*}p} \propto W^{2\lambda}$. 
For $Q^2 \ge 1.5$ GeV$^2$, the power $\lambda$ is between 0.15 and 0.35
\cite{f2_zeus,f2_h1}. In contrast, the total cross-section for 
real photon-proton scattering (photoproduction, with $Q^2 = 0$) 
shows only a modest rise with $W$, 
$\lambda=0.08$ \cite{DL-2}, consistent with the energy behaviour 
of the total $p\overline{p}$ cross-section.  Regge theory has 
been used successfully by, {\em e.g.,} Donnachie and Landshoff 
(DL) \cite{DL-2} to describe the energy dependence of the total 
cross-section for hadron-hadron scattering and real photon-proton
scattering, but their prediction \cite{DL} for virtual 
photon-proton scattering fails to describe the data for $Q^2 \ge 1.5$ 
GeV$^2$.  Different groups
(CKMT\cite{CKMT}, BK\cite{BK}, ABY\cite{ref-ABY}, and ALLM\cite{allm}) 
have used a variety of approaches to connect 
the very low $Q^2$ behaviour with high $Q^2$ pQCD.  CKMT extend the 
Regge prediction by 
including $Q^2$ dependent absorptive corrections that modify the 
effective pomeron intercept, resulting in a $Q^2$ dependent $\lambda$
up to some $Q^2_0$ in the range 1 to 5 GeV$^2$; pQCD 
is then used to evolve from this $Q_0^2$ to higher $Q^2$. BK 
describe the $Q^2$ behaviour using a generalised vector dominance
model (GVDM): the low $Q^2$ region is controlled by the 
contributions of the low mass vector mesons, and the higher mass 
contributions are adjusted to provide agreement with pQCD 
predictions using a standard set of structure function 
parametrisations at larger $Q^2$. 
ABY extend their high $Q^2$ QCD-inspired parametrisation into the 
low $Q^2$ 
regime, and modify the evolution of $\alpha_s$ so that it 
saturates at a finite value.
ALLM introduce
parametrisations that interpolate between the Regge and pQCD 
regimes. A review is given in reference \cite{ref-Levy}.  

To study the transition from the hadronic type behaviour at
$Q^2 \approx 0$ to the deep inelastic scattering (DIS) regime 
($Q^2 \gg 1$ GeV$^2$), the kinematic coverage of the ZEUS 
detector was substantially extended starting in 1995 with the 
installation of a new beampipe calorimeter (BPC)\footnote{The new BPC 
replaced a previous calorimeter
described in \cite{martin}.}. 
Here we report on the measurement of $F_2$ and $\sigma_{tot}^{\gamma^{*}p} $ for 
$0.11 \le Q^{2} \le 0.65$ GeV$^2$ from $e^+p$ scattering at 
$\sqrt{s}=300$ GeV using the ZEUS detector with the new 
BPC.  This analysis is based on 1.65 pb$^{-1}$ 
of data taken during the 1995 HERA run. Recently
the H1 Collaboration has reported an $F_2$ measurement in four bins 
in the $Q^2$ range of 0.35 to  0.65 GeV$^2$ in a somewhat 
different $W$ range \cite{h1-svtx}. For $Q^2 \ge$ 0.23 GeV$^2$,
the E665 Collaboration reported  a measurement of the proton structure 
function  at much higher $x$\cite{e665} than this analysis.

\section{Kinematics}
Inelastic positron-proton scattering, 
\begin{equation}
e^+p\rightarrow e^+ X 
\label{eq:ep}
\end{equation}
can be described in terms of two kinematic variables, $x$ and
$Q^2$, where $x$ is the Bjorken scaling variable and $Q^2$ the 
negative of the
square of the four-momentum transfer. In the absence of 
initial and final state radiation, $Q^2=-q^2=-(k-k^{\prime})^2$ 
and $x=Q^2/(2P\cdot q)$, 
where $k$ and $P$ are the four-momenta of the incoming positron 
and proton respectively and $k^{\prime}$ is the four-momentum of 
the scattered positron. The fractional energy transfer to the 
proton in its rest frame, $y$, can be related to $x$ and $Q^2$ by
$y=Q^2/(sx)$, where $s$ is the square of the $e^+p$ c.m.  
energy which is given by $s\simeq 4E_eE_p$. Here, $E_e$ (27.5 GeV) and 
$E_p$ (820 GeV) are the positron and proton beam energies, respectively. 
The kinematic variables, $y$ and $Q^2$, are related to the 
energy, $E^\prime_e$, and angle with respect to the proton beam direction, 
$\theta_e$,  of the scattered
positron.  We also use $\vartheta = \pi - \theta_e$, the 
angle with respect to the positron beam direction. Scattering
at low $Q^2$ results in positrons emerging at small 
$\vartheta$,

\begin{equation}
y=1-\frac{E_e'}{2E_e}(1+\cos \vartheta)\approx
1-\frac{E_e'}{E_e},\,\,\,\,Q^2=
2E_eE_e'(1-\cos\vartheta)\approx E_eE_e' \vartheta^2. 
\label{eqn-emethod}
\end{equation}

\section{Experimental setup and kinematic reconstruction}

The ZEUS detector  \cite{zdet} is a general purpose magnetic detector at 
the HERA collider. To enhance the 
acceptance of the detector at small $Q^2$, two beam pipe calorimeter
modules (BPC)  \cite{BPCNIM} located on two sides of the beam pipe
at $2937$ mm from the interaction point in the rear 
(positron) direction\footnote{The ZEUS right-handed coordinate system is 
defined with the origin at the nominal interaction point
by the Z axis pointing in the proton beam direction and the X axis 
horizontally pointing towards the center of HERA.}
were installed, as shown in figure \ref{fig:uniformity}(a). 
The BPC covers positron 
scattering angles relative to the incident direction of 15 to 34 mrad. 
At these small angles, the maximum possible scattered positron energy is
equal to the beam energy, 27.5 GeV. The beam pipe has two low-mass 
windows (0.016 radiation length (r.l.)) in front of the BPC to allow 
positrons to exit the beam pipe with minimal interference.

The BPC is an electromagnetic scintillator sampling calorimeter. 
The passive absorber layers consist of twenty-six 13.8 cm $\times$  
13 cm $\times$ 0.92 r.l. thick tungsten alloy plates;
the active layers consist of  7.9 mm wide and 2.6 mm thick 
scintillator strips alternating each layer in the horizontal and 
vertical directions. 
The scintillator strips are read out from one end using wavelength 
shifters (WLS). Each WLS is coupled to  a miniature photo-multiplier 
tube (Hamamatsu R5600-03). 
The vertically oriented scintillator strips provide the X  position 
measurement and the horizontally oriented strips the Y measurement.  
The readout electronics are similar to those used for the main ZEUS
uranium scintillator calorimeter (CAL)  \cite{zdet}. 
The alignment is known to an accuracy of 0.5 mm from an optical 
geometrical survey. The distance between the two calorimeter 
modules on either side of the beam pipe is mechanically 
constrained to within 0.2 mm. Due to synchrotron radiation from 
the positron beam, the modules are placed asymmetrically around 
the beam. Only one of the two BPC modules is used for physics 
analysis due to the very small acceptance of the other module, 
which is used exclusively for alignment purposes. The typical 
geometrical acceptance is 10\%.

\subsection{Detector simulation and response}
\label{e_scale}

The BPC simulation is based on the GEANT  \cite{mc_geant} program, with an 
independent check performed using EGS4  \cite{egs4}. The energy 
spectra for 
1 to 6  GeV incident electrons in the simulation are in good agreement 
with test beam data 
taken at these energies,  and are consistent with the design energy 
resolution of $17\%/\sqrt{E}$. The non-linearity is found to be less than 1\%
in the simulation for 2 to 6 GeV incident electrons, in agreement with test
beam data.

The energy calibration was performed {\em in situ} using {\em 
kinematic peak} (KP) events\footnote{A  
cut $y_{JB} <0.04$ (see below) selects scattered positrons whose 
energy distribution sharply peaks within 2\% of  the beam energy,
providing a good calibration source  \cite{zeus-KP}.} in two steps:
a relative strip-to-strip calibration, followed by
an overall energy scale calibration.
Figure ~\ref{fig:uniformity}(b) shows the fractional deviation of
the KP energy from the mean value as a function of the scattered
positron X impact position at the BPC after the relative 
calibration. The energy scale is uniform to within 0.5\% over the
BPC fiducial region, which extends to 8 mm from the edge of the 
BPC. The overall energy scale was established by comparing the KP
energy spectra of data and simulation, which included QED 
radiative corrections. A $\chi^2$ was calculated between the two 
spectra, and the energy scale of the data adjusted to minimise 
it.

The absolute energy scale obtained with KP events was checked 
using elastic $\rho^0$ events, \linebreak $e^{+}p\rightarrow 
e^{+}\rho^0p$. The scattered positron energy and 
position were measured in the BPC, and the three-momenta of the 
two $\rho^0$ decay pions were measured using the ZEUS Central 
Tracking Detector (CTD)\cite{ctd}.  Using the four-momentum of the $\rho^0$,
the scattered positron energy can be independently determined. 
Figure ~\ref{fig:uniformity}(c) shows the ratio of the measured 
positron energy to that determined from the CTD, for both 
simulation and data. Radiative corrections are responsible for
the tail at low values and the fact that the distributions peak below unity.
The agreement between the data and the simulation is very good. Consequently, 
it was concluded that the energy scale determined using the KP events was 
accurate to within 0.5\%, and that the BPC resolution was well modeled
in the simulation.

The BPC was located only 4.4 cm from the beam in the horizontal 
direction, and received 12 kGy of radiation during the 1995 HERA 
running period. The resulting 
damage caused a drop in the energy scale of up to 2.5\% for the 
regions closest to the beam, determined using KP events. To 
correct for this, the data were separated into 4 time periods and
the energy calibration performed separately for each period.
The amount of radiation damage to the BPC was also determined 
using a movable cobalt source scan calibration 
system  \cite{CAL_cobalt} and by measuring the response of a sample
of scintillator strips after disassembling the calorimeter at the
end of 1995 after data taking.  The degradation of each 
individual scintillator strip was built into the EGS4 simulation,
which showed that
the non-linearity from 10 GeV to full scale due to radiation 
damage was less than 1\%.

The BPC measured the arrival time of the positron with an accuracy 
determined to be 0.4 ns for positrons with an energy
greater than 6 GeV.

\subsection{Positron identification and position reconstruction}
\label{sect-shower}
The scattered positron position in the BPC was determined using 
the logarithmic energy-weighted shower position  \cite{pos_rec} 
using scintillator strips containing more than 4\% of the total 
shower energy. A RMS resolution of 1.3 mm with a maximum systematic shift of 
0.5 mm was obtained for 5 GeV incident electrons in the 
beam test. This was well-reproduced in the simulation. The 
position resolution improves gradually with increasing positron 
energy. The resolution was determined from the simulation to be 
0.6 mm at 27.5 GeV for positron impact positions within the fiducial region of
the BPC. 

As a cross-check of the absolute position, QED Compton events 
$e^{+}p \rightarrow e^{+}\gamma p$ were used. Since both the photon and the scattered
positron are detected in the BPC modules, these events provide an 
over-constrained kinematic system.  Using the accurately known
distance between the two BPC modules and the QED Compton event 
kinematics, the position of each BPC module relative to the 
positron beam was determined. The agreement with the survey is 
better than 0.5 mm. The accuracy of the QED Compton method is 
dominated by the uncertainty of the calorimeter energy scale: 0.5\%, 
corresponding to 0.5 mm.

Positron identification was performed using the transverse size of the
shower. The second moments of the shower in the X and Y directions, 
$\sigma_X$ and  $\sigma_Y$, were calculated using the logarithmic
energy weighted method mentioned above.
The combination 
$\sqrt{(\sigma_X^2+\sigma_Y^2)/2}$ was required to be
less than 7.5 mm. This yielded a positron acceptance in excess of
95\% at 7 GeV, rising to 98\% above 12 GeV,  while rejecting 
hadrons and positrons that have preshowered in the beam-pipe 
wall, which have a much wider transverse width. Using a  
sample of KP positrons, the transverse energy profile in the 
simulation was tuned to that of the data. 
\subsection{Vertex  determination and luminosity measurement}

The position of the $e^+p$ interaction vertex is needed to determine the 
positron scattering angle. The Z position of the vertex was measured 
using the CTD on an event-by-event basis with a typical resolution of 4 mm.  
For events  with no CTD vertex 
information (about 8\% of the total), the Z position of the vertex was 
set to the average position of the full data sample, $<$Z$>$ = $-3$ cm. 
The longitudinal size of the luminous region was about 12 cm (r.m.s.).
The mean values of the X and Y vertex positions, determined on a run-by-run 
basis, were used. The transverse sizes of the beam in X and
Y were about 300 $\mu m$ and 70 $\mu m$, respectively.

The luminosity was measured via the bremsstrahlung process 
$e^+p\rightarrow e^+\gamma p$, using a separate 
electromagnetic calorimeter detector system 
(LUMI)  \cite{lumi}.  A lead-scintillator calorimeter 
positioned at Z $=-$107 m, accepting photons with scattering angles 
less than 
0.5 mrad, was used for the luminosity measurement. 
The uncertainty of the 
luminosity measurement for the data sample used in this analysis is 1.5\%. 
A second 
electromagnetic calorimeter, positioned at Z $=-$35 m, was used for tagging 
positrons in background studies. 

\subsection{Reconstruction of the kinematic variables} 

In this analysis, the kinematic variables $y$ and $Q^2$ are reconstructed
using the energy and angle of the scattered positron, determined using the BPC
and the CTD vertex position with equation~\ref{eqn-emethod}.
Using this method of reconstruction (``electron method''), $y$ and $Q^2$
can be determined reliably over the kinematic range $y > 0.1$ and 
$Q^2 > 0.1$ GeV$^2$. The $y$ resolution is 0.02 to 0.04, and 
the $Q^2$ resolution is 6 to 8\%. 

As the ZEUS CAL is an almost hermetic detector, it can be used 
to measure the hadronic system, denoted by $X$ in equation 
\ref{eq:ep}. The following quantities are reconstructed using 
the CAL,

\begin{equation} 
\delta_{CAL} = \sum_iE_i(1-\cos \theta_i),\,\,\,
y_{JB} = \frac{\delta_{CAL}}{2E_e},\,\,\, E_{tot} =\sum_i E_i \label{eq:yjb}, 
\end{equation}

where $E_i$ is the energy measured in the $i^{th}$ CAL cell and $\theta_i$ is the 
polar angle of the center of the $i^{th}$ calorimeter cell with respect to the 
positive Z axis; the sum extends over all cells in the CAL.  The quantity 
$y_{JB}$ provides a measure of the kinematic variable $y$, and has superior
resolution at very low $y$ compared to that from the electron method. 
$E_{tot}$ is the measured total energy of the
hadronic system. These quantities were used in the trigger and event 
selection for rejecting background, reducing QED radiative 
corrections and controlling event migration at low $y$. 

\section{Trigger, data taking and event selection}
    
ZEUS selects events online using a three-level trigger 
system   \cite{z_trigger}.
Both the energy and timing information from the BPC were used for 
the First Level Trigger (FLT). An energy cut of 6 GeV was
made and the timing was required to be consistent with an $e^+p$ collision.
Proton-gas events occurring upstream were also rejected by timing 
measurements made by scintillation 
counter arrays situated along the beamline at $Z=-730, 
-315$, and $-150$~cm, respectively. For the Second Level Trigger, 
CAL timing information was used to reject non $e^+p$ 
events. An approximate value of $y_{JB}$, determined from the CAL
energies assuming an interaction vertex at Z = 0, was required to
be greater than 0.02 and the total CAL energy was required to be 
greater than 3 GeV. No additional BPC cuts were imposed at the 
Third Level Trigger.

The FLT efficiency was studied using a sample 
of events triggered only by the CAL.   
All of the offline event selection cuts (see below) were applied to this 
sample. The trigger was found to be fully efficient for BPC 
energies greater than 7 GeV, as shown in 
figure~\ref{fig:uniformity}(d).

The offline event selection cuts are as follows.  
The reconstructed positron is required to be  
within the BPC fiducial region, to have more than 7 GeV energy, 
and to pass the shower width cut described in section 
\ref{sect-shower}. The BPC time is required to be within 3 ns of
the time for $e^+p$ interactions. If the event vertex is 
well reconstructed with the CTD, the Z position of the vertex is
required to be within the window -40 cm $<$ Z$_{VTX} < 100$ cm. The
quantity $\delta = \delta_{CAL} + 2E^\prime_e$, is required to 
lie in the range 35 to 60 GeV; $\delta$ equals twice the positron
beam energy (55 GeV) for a completely contained event, but the 
distribution peaks at much lower values for photoproduction 
events where the scattered positron is lost in the rear 
direction. A cut $y_{JB} > $0.06 reduces migration of events from
very low $y$, where the resolution of the electron method is 
poor. Finally, if timing information from the CAL is 
available, the time  is required to be consistent with an $e^+p$ collision. 
After cuts, 109105 events remain in the data sample. 

\section{Analysis}

\subsection{Physics simulation}
\label{sect-physicssimulation}
A physics simulation is used to determine the 
radiative corrections and the acceptance of the detector.   
The starting point for the 
simulation of $e^+p$ collisions in the $y$ and $Q^2$ region of this
measurement is the program DJANGO 6.22, which interfaces the 
programs HERACLES  \cite{mc_her} 4.5.1, ARIADNE  \cite{mc_ari} 4.06 
and LEPTO  \cite{mc_lep} 6.4.1. The HERACLES program calculates the
structure functions $F_2$ and $F_L$ from an input set of parton 
density functions. From these, it calculates the differential 
cross-section including initial and final state radiation, and 
the full one-loop virtual corrections. ARIADNE implements the 
colour dipole model for gluon radiation between the struck quark 
and the proton remnant. Finally, LEPTO handles the fragmentation 
using the program JETSET  \cite{mc_jet}.

Several modifications were made to the program to generate events at 
low $Q^2$. 
(1) The Donnachie-Landshoff parameterisation  \cite{DL} was used to 
calculate $F_2$ in the $Q^2$ range of this measurement.
\linebreak
(2) The longitudinal structure function $F_L$ was set equal to zero at 
this stage. (3) Elastic vector meson (VM) $\rho^0$ 
events were generated. The VM events were re-weighted according 
to the cross-section and $W$ dependence recently measured at ZEUS
using the BPC. This contribution amounted to 6\%
of the events  \cite{warsaw_rho}. (4) Diffractive events were 
generated according to $d^2\sigma/dtdM^2_X \sim e^{bt}/(M^2_X +
Q^2 - M^2_{\rho})^{1.1}$  \cite{epsoft}, where $t$ is the square of
the four-momentum transferred to the outgoing proton, $M_X$ is the 
invariant mass of the hadronic final state, and $b=6$ 
GeV$^{-2}$. The fraction of diffractive events (around 25\%) was 
determined from the data by counting the events with a 
characteristic rapidity gap; that is, a region of little or no 
hadronic activity between the forward edge of the CAL and the jet from 
the struck quark.  The overall acceptance is weakly sensitive to 
changes in the relative contributions of the different event 
types (see section 5.5).

A generated event sample in the region $Q^2>$ 0.05 GeV$^2$ and 
$y>0.03$ corresponding to more than twice the 
luminosity of the data was passed through the complete ZEUS 
simulation chain, which is based on the GEANT  \cite{mc_geant} 
program, and then processed using the same offline reconstruction
software as for the data.  The general characteristics of 
the data are well-described by the simulation, as shown in figure
\ref{fig:empz}(a-c).

\subsection{Binning of the data}

The data are binned in the variables $y$ and $Q^2$, which 
makes efficient use of the available phase space, as the lower electron 
energy cut corresponds to an upper $y$ cut. 
The accessible kinematic region lies between $y$ values of 0.08 and 0.74, and
$Q^2$ values of 0.1 and 0.74 GeV$^2$. The sizes of the bins are 
chosen based on the experimental resolution of the
kinematic variables and the number of events. At low $y$, the bin
widths in $y$ are chosen to be twice the $y$ resolution;  for $y$
greater than 0.37, bins of approximately constant width are used.
The lowest $Q^2$ bin has a width 2.5 times the $Q^2$ resolution; 
higher $Q^2$ bins have a constant width in $\log(Q^2)$ which 
results in approximately constant numbers of events in each $Q^2$
interval.

For positrons within the BPC fiducial region, the efficiency of 
the event selection cuts is close to 100\% for $y < 0.5$, 
decreasing to about 70\% at $y=0.7$.  The purity, defined as the 
fraction of events reconstructed in a bin that were generated in 
that bin, is typically 50\%.

\subsection{Background determination}

The background from beam-gas interactions is determined using 
unpaired positron and proton bunches. The size of this background
is 1\%, and is subtracted statistically. The dominant
source of background comes from photoproduction interactions 
where the scattered positron escapes through the rear beam pipe 
and a fake positron is reconstructed in the BPC.  Typically such 
events have much lower $\delta$ values than signal events. 
Photoproduction background events were generated using the PYTHIA
program  \cite{pythia} with the ALLM cross-section parameterisation
 \cite{allm}. This PYTHIA sample is used to perform a bin by bin 
subtraction of the photoproduction background.
Figure \ref{fig:empz}(d) shows the $\delta$ spectra for data (solid circles),
signal simulation (dashed line), photoproduction simulation 
(shaded region), and the sum of the signal and photoproduction simulations 
(solid line). The good agreement between this sum and the data suggests that 
the photoproduction background is well simulated, and gives  
a contamination of a few percent in most bins, rising up to 15\% in the
highest $y$ bins.

As a cross-check, use was made of the very small angle LUMI 
positron detector to measure the photoproduction background 
directly. This detector accepts scattered positrons with 
$Q^2<0.01$ GeV$^2$ and $0.2<y<0.6$ and may be used to tag events 
with a fake BPC positron signal that pass the event selection 
cuts. The measured background is shown as the triangular points 
in figure \ref{fig:empz}(d); the $y<0.6$ cut limits the 
measurement to $\delta<35$ GeV. Once again, the measured points 
are in good agreement with the simulated background.

The contamination from higher $Q^2$ DIS events is less that 0.1\% based on
searches in the CAL for additional positron candidates in both data
and  the simulation.
\subsection{Determination of {\boldmath $F_2$} and 
{\boldmath $\sigma_{tot}^{\gamma^{*}p}$}}

The double-differential $e^+p$ cross-section for inelastic 
scattering can be expressed in terms of 
the total cross-section for virtual transverse (T) and longitudinal (L) 
photons:

\begin{equation}
\frac{d^2\sigma(ep \rightarrow eX)}{dydQ^2}
=  \Gamma \left[ \sigma_T(y,Q^2) + \epsilon \sigma_L(y,Q^2)\right]
(1+\delta_r(y,Q^2)) 
\label{eqn:dsigma}
\end{equation}

where the flux $\Gamma \approx \alpha (1+(1-y)^2)/(2\pi Q^2 y)$, the 
fractional flux of longitudinally polarised virtual 
photons $\epsilon \approx 2(1-y)/(1+(1-y)^2)$, and
$\delta_r$ is the radiative correction factor. These cross-sections can be 
related to the proton structure functions $F_2$ and $F_L$ 
by $F_2=(Q^2+4m_p^2x^2)(\sigma_T+\sigma_L)/(4\pi^2\alpha(1-x))$,
$F_L=(Q^2+4m_p^2x^2)\sigma_L/(4\pi^2\alpha(1-x))$ and the total virtual 
photon-proton cross section by
$\sigma_{tot}^{\gamma^{*}p}\equiv\sigma_T+\sigma_L$. In the $Q^2$
range of this analysis, the contribution from $Z^0$ exchange is 
negligible. The radiative correction to the Born cross-section, 
$\delta_r$, is a function of 
$y$ and $Q^2$, but to a good  approximation is independent of $F_2$. 

An iterative procedure is adopted to extract the sum $\sigma_T + 
\epsilon \sigma_L$. Monte Carlo events are generated as 
described in section \ref{sect-physicssimulation} to determine 
the acceptance and efficiency of the cuts in each bin. The first
$\sigma_T + \epsilon \sigma_L$ values are then fit   
with a simple functional form, inspired by the ALLM 
parameterisation~\cite{allm}. The result of the fit is used to 
reweight the input distributions in the simulation event by event
to re-evaluate the acceptance and efficiency.
New values of $\sigma_T + \epsilon \sigma_L$ in each bin are 
calculated and the procedure repeated until the change between subsequent
iterations is less than 0.5\% in all bins; three 
iterations are required.  The relative fractions of diffractive 
and VM events are kept fixed in this procedure.

Once the $\sigma_T + \epsilon \sigma_L$ values are determined,  $F_2$ and 
$\sigma_{tot}^{\gamma^* p}$ are calculated assuming $\sigma_L$ 
to be either zero or the value given by the Vector Dominance Model 
(VDM), $\sigma_L=K(Q^2/M_V^2)\sigma_T$. $M_V$ is set equal to 
the mass of the $\rho^0$ (0.77 GeV) and $K$ to 0.5.

\subsection{Systematic uncertainties}
\label{sec-systematics}

To estimate the systematic uncertainties of the measured $F_2$ values, the 
following checks were performed. In each case, an aspect of the event
selection, reconstruction of kinematic variables, 
or $F_2$ determination was modified, the procedure described above was 
repeated and the change in $F_2$ noted.
(1) The 0.5\% uncertainty on the energy scale (see Section~\ref{e_scale}) gave
an effect of approximately 3\% for $F_2$; 
the non-linearity of the BPC, estimated to 
be less than 1\%, resulted in changes of less than 5\%.
(2) Varying the absolute position by 0.5 mm produced changes of less than 6\%.
(3) To estimate the uncertainty due to the electron finding efficiency, 
the shower width cut was varied by 1 mm yielding a change up to 2\% at high 
$y$. 
(4) The uncertainty due to the CAL event selection cuts was 
checked by varying the CAL energy scale by 3\%, changing 
the CAL minimum cell energy threshold, varying the $\delta$ cut by 2 GeV 
and varying the $y_{JB}$ cut by 0.01. The effect on 
$F_2$ is small for moderate $y$ bins, and reaches 6\% at high and low $y$. 
(5) The estimated uncertainty on the photoproduction 
background was 30\%, obtained from a comparison of the 
various methods described above. Consequently, the amount of 
background subtracted was varied by the same amount. The effect 
on $F_2$ increases with $y$, reaching 4\% in the highest $y$ bins. 
(6) The fraction of diffractive and VM events was varied by 25\%. The
effect on $F_2$ was small except for the lowest $y$ points where it 
reached 4\%.
(7) Uncertainties due to the description of the
hadronic final state in the simulation were estimated by comparing the 
results from different simulation programs(EPSOFT  \cite{epsoft}, 
PYTHIA  \cite{pythia}, HERWIG{  \cite{herwig}); constraints were 
provided by comparing hadronic distributions measured in the data
with those predicted by the simulations. An error of 2\% in $F_2$
for $y$ less than 0.4, rising linearly to 5\% at $y$ = 0.74, was 
assigned. (8) The uncertainties in the radiative corrections, which
modify the Born cross-section by  10-15\%, were  investigated.
As mentioned in section \ref{e_scale}, the 
tail of the $E_{BPC}/E_{calc}$ distribution for 
elastic $\rho^0$
events (figure 1(c)) is due to initial state radiation. Comparison of 
the simulation with data in that region provides an estimate of 
the uncertainty in the radiative correction.  This uncertainty represents 
possible changes of 3 to 4\% on $F_2$.

The extracted uncertainties after each of the above systematic checks
are displayed in table \ref{tab-systematics} (see Appendix).
The total systematic error for each bin was determined by adding 
the changes to $F_2$ from different checks in quadrature. 
These are shown in table~\ref{tab:f2}.
The systematic errors are around
6\% for moderate $y$ bins and are dominated by BPC calibration and 
radiative correction uncertainties. 
For low $y$, uncertainties in the CAL energy scale, as 
well as the dependence of the acceptance on the fraction of 
diffractive events leads to uncertainties up to 11\%. At high
$y$, uncertainties on the contributions from photoproduction 
background and the properties of the hadronic final state 
resulted in errors as high as 8\%.

The uncertainties in the luminosity measurement and trigger efficiency 
contribute to the overall normalisation error, which amounts to 
2.4\%.  The statistical errors are in the range 2--4\%.

\section{Results}
\begin{table}[htb]
\begin{center}
\begin{tabular}{|c|c|c|c||c|c|c|c|}
\hline
\rule[-2mm]{0mm}{6mm}$Q^2$ & $y$ & $F_2$ & $\Delta F_2[F_L]$ &
$Q^2$  & $y$ & $F_2$ & $\Delta F_2[F_L]$ \\ \hline
\rule[-2mm]{0mm}{6mm}0.11 & 0.60 & 0.163$\pm 0.005 \pm 0.011$ & 0.005 &
0.25 & 0.60 & 0.267$\pm 0.010 \pm 0.020$ & 0.017 \\ \hline
\rule[-2mm]{0mm}{6mm}0.11 & 0.70 & 0.174$\pm 0.006 \pm 0.014$ & 0.008 &
0.30 & 0.12 & 0.263$\pm 0.005 \pm 0.022$ & 0.000 \\ \hline
\rule[-2mm]{0mm}{6mm}0.15 & 0.40 & 0.188$\pm 0.006 \pm 0.015$ & 0.003 &
0.30 & 0.20 & 0.280$\pm 0.005 \pm 0.014$ & 0.002 \\ \hline
\rule[-2mm]{0mm}{6mm}0.15 & 0.50 & 0.203$\pm 0.005 \pm 0.013$ & 0.005 &
0.30 & 0.26 & 0.295$\pm 0.007 \pm 0.017$ & 0.003 \\ \hline
\rule[-2mm]{0mm}{6mm}0.15 & 0.60 & 0.200$\pm 0.006 \pm 0.014$ & 0.008 &
0.30 & 0.33 & 0.296$\pm 0.008 \pm 0.019$ & 0.005 \\ \hline
\rule[-2mm]{0mm}{6mm}0.15 & 0.70 & 0.205$\pm 0.008 \pm 0.018$ & 0.012 & 
0.30 & 0.40 & 0.301$\pm 0.010 \pm 0.018$ & 0.008 \\ \hline
\rule[-2mm]{0mm}{6mm}0.20 & 0.26 & 0.225$\pm 0.005 \pm 0.016$ & 0.002 &
0.30 & 0.50 & 0.305$\pm 0.011 \pm 0.021$ & 0.014 \\ \hline
\rule[-2mm]{0mm}{6mm}0.20 & 0.33 & 0.227$\pm 0.006 \pm 0.011$ & 0.003 &
0.40 & 0.12 & 0.332$\pm 0.007 \pm 0.027$ & 0.001 \\ \hline
\rule[-2mm]{0mm}{6mm}0.20 & 0.40 & 0.231$\pm 0.005 \pm 0.013$ & 0.005 & 
0.40 & 0.20 & 0.337$\pm 0.008 \pm 0.020$ & 0.002 \\ \hline
\rule[-2mm]{0mm}{6mm}0.20 & 0.50 & 0.238$\pm 0.006 \pm 0.015$ & 0.008 &
0.40 & 0.26 & 0.367$\pm 0.010 \pm 0.020$ & 0.005 \\ \hline
\rule[-2mm]{0mm}{6mm}0.20 & 0.60 & 0.254$\pm 0.009 \pm 0.017$ & 0.013 & 
0.40 & 0.33 & 0.358$\pm 0.012 \pm 0.020$ & 0.008 \\ \hline
\rule[-2mm]{0mm}{6mm}0.20 & 0.70 & 0.257$\pm 0.011 \pm 0.021$ & 0.019 & 
0.40 & 0.40 & 0.392$\pm 0.014 \pm 0.024$ & 0.013 \\ \hline
\rule[-2mm]{0mm}{6mm}0.25 & 0.20 & 0.249$\pm 0.005 \pm 0.012$ & 0.001 & 
0.50 & 0.12 & 0.351$\pm 0.009 \pm 0.027$ & 0.001 \\ \hline
\rule[-2mm]{0mm}{6mm}0.25 & 0.26 & 0.256$\pm 0.005 \pm 0.013$ & 0.002 & 
0.50 & 0.20 & 0.375$\pm 0.010 \pm 0.019$ & 0.003 \\ \hline
\rule[-2mm]{0mm}{6mm}0.25 & 0.33 & 0.264$\pm 0.006 \pm 0.015$ & 0.004 & 
0.50 & 0.26 & 0.414$\pm 0.013 \pm 0.021$ & 0.006 \\ \hline
\rule[-2mm]{0mm}{6mm}0.25 & 0.40 & 0.276$\pm 0.007 \pm 0.019$ & 0.006 & 
0.65 & 0.12 & 0.386$\pm 0.012 \pm 0.039$ & 0.001 \\ \hline
            0.25 & 0.50 & 0.275$\pm 0.009 \pm 0.018$ & 0.011 & 
0.65 & 0.20 & 0.464$\pm 0.018 \pm 0.022$ & 0.004 \\ \hline
\end{tabular}
\caption{Measured $F_2$ values, with the assumption $F_L=0$.  The
first error indicates the statistical error, and the second 
indicates the systematic error.  The values in the column marked 
$\Delta F_2[F_L]$ show the change in $F_2$ assuming a value for 
$F_L$ given by VDM (see text), and are not included in
the systematic errors. The bin boundaries in $y$ are
0.08, 0.16, 0.23, 0.30, 0.37, 0.45, 0.54, 0.64, and 0.74 and in $Q^2$
0.1, 0.13, 0.17, 0.21, 0.27, 0.35, 0.45, 0.58, and 0.74 GeV$^2$.
} \label{tab:f2} \end{center} \end{table}

In the following, $F_2$ is presented as a function of $x$ (or 
$y$) and $Q^2$, and $\sigma_{tot}^{\gamma^*p}$ as a function of 
$W^2$ and $Q^2$. While the former vanishes as $Q^2 \rightarrow 
0$, the latter is expected to extrapolate smoothly to the total 
photoproduction cross-section.

The $F_2$ results are displayed in table~\ref{tab:f2} in bins 
of $Q^2$ and $y$ and plotted in figure~\ref{fig:f2_x_theory} as a 
function of $x$ for different Q$^2$ bins. Here $F_L$ is 
assumed to be zero. Assuming $F_L$ as given by VDM, the effect on
$F_2$ is typically around 1-2\% for most bins, and increases 
$F_2$ by up to 7\% in the lowest $x$ bins, as shown in table 
\ref{tab:f2}.

Also shown in figure \ref{fig:f2_x_theory} are the data from the
E665\cite{e665} experiment at similar $Q^2$, but much larger $x$ 
values, and four recent points from the H1\cite{h1-svtx} 
experiment with $Q^2\le 0.65$ GeV$^2$. One observes a rise of 
$F_2$ by a factor 1.5 to 2 from $x$ near $10^{-3}$ to $x$ around 
$10^{-5}$. In the bottom of figure~\ref{fig:f2_x_theory}, the 
$F_2$ values for $Q^2=$1.5, 3.0 and 6.5 GeV$^2$ are shown, taken 
from previous H1 and ZEUS publications \cite{f2_zeus, f2_h1, 
h1-svtx} and from E665. They illustrate the rapid rise of $F_2$ 
with decreasing $x$ as $Q^2$ increases.

Curves from various theoretical models are overlaid. In general DL and CKMT 
predictions are
15\% and 10\% lower than the data respectively, while the BK prediction 
is 15\% higher. No attempt was made to modify the parameters of these models 
to fit the data. The value of $F_2$ given by GRV 
is approximately 35\% of the measured 
value at $Q^2=0.4$ GeV$^2$, rising to about 80\% at $Q^2=0.65$ 
GeV$^2$. At larger $Q^2$ the GRV predictions reproduce the 
rapid rise of $F_2$, but tend to lie somewhat above the data.  
The ABY parametrisation, which included preliminary ZEUS BPC $F_2$ 
results in the parameter fit, gives a good description of the 
data.

Figure~\ref{fig:sig_w2} shows $\sigma_{tot}^{\gamma^*p}$ at low 
$Q^2$ and at higher $Q^2$ taken from previous 
ZEUS analyses and from the H1 and E665 
experiments as a function of $W^2$. The total cross-section for 
real $\gamma p$ scattering is also shown  \cite{zgp,hgp,low_w}. 
The curves in figure~\ref{fig:sig_w2} show the expectations from 
the soft Pomeron model of DL (dotted curve) and the pQCD model of
GRV (dashed curve). The $W$ dependence of the DL predictions is
given by $\sigma_{tot}^{\gamma^*p} \propto W^{2\lambda}$ with 
$\lambda=0.08$ independent of $Q^2$. The GRV model predicts a 
stronger variation with $Q^2$ and $W^2$. Although the DL model 
curve is below the ZEUS data, its  slope is in broad 
agreement for $Q^2\le 0.65$ GeV$^2$; however, this is not true 
for $Q^2\ge 1.5$ GeV$^2$.  For $Q^2\ge 1.5$ GeV$^2$, the GRV 
model produces a rapid rise of $\sigma_{tot}^{\gamma^*p}$ with 
$W^2$ that is in better agreement with the data.

\section{Conclusions}
In previous studies, ZEUS and H1 have shown that for $Q^2 \ge 1.5$ GeV$^2$ 
and $x \ll 10^{-2}$, the proton structure function $F_2$ rises rapidly as 
$x$ decreases, in agreement with models based on 
perturbative QCD. In this paper we have presented a
measurement of $F_2$  and $\sigma_{tot}^{\gamma^*p}$ for  
$0.11\le Q^2\le 0.65$ GeV$^2$ and $2 \times 10^{-6} \le x \le  6 
\times 10^{-5}$ ($100 < W < 230$ GeV), covering the 
region between deep inelastic scattering 
and photoproduction.  
Similar results have recently been reported by 
H1\cite{h1-svtx}. In 
combination with data from E665 for $x \ge 10^{-3}$, $F_2$ 
exhibits a modest rise in this low $Q^2$ region. Together with 
previously published HERA data for $Q^2 \ge 1.5$ GeV$^2$, where the 
rise is more rapid, our results suggest that pQCD calculations 
can account for a significant fraction of the cross-section 
starting at $Q^2\approx 1$ GeV$^2$.

\section*{Acknowledgements}
We thank the DESY Directorate for their strong support and encouragement
and the diligent efforts of the HERA machine group. We are also  grateful for
the support of the DESY computing and network services.
The design, construction,
and installation of the ZEUS detector has been made possible by the ingenuity
and dedicated effort of many people from DESY and the home institutes
who are not listed as authors. In particular, we would like to thank 
M.~ Gospic, H.~ Gr\"{o}nstege, J.~ Hauschildt,  K.~ L\"{o}ffler,  M.~ Riera, 
H.~ Schult, W.~ Sippach, E.~ Weiss for their 
important contributions to the construction and installation of 
the BPC.

\begin{figure}[t]
\setlength{\unitlength}{\textwidth}
\begin{picture} (1.0,1.0)
\put (0.35,0.975){\mbox{\Huge\bf ZEUS 1995}}
\put (0,0.5){\mbox{\epsfig{file=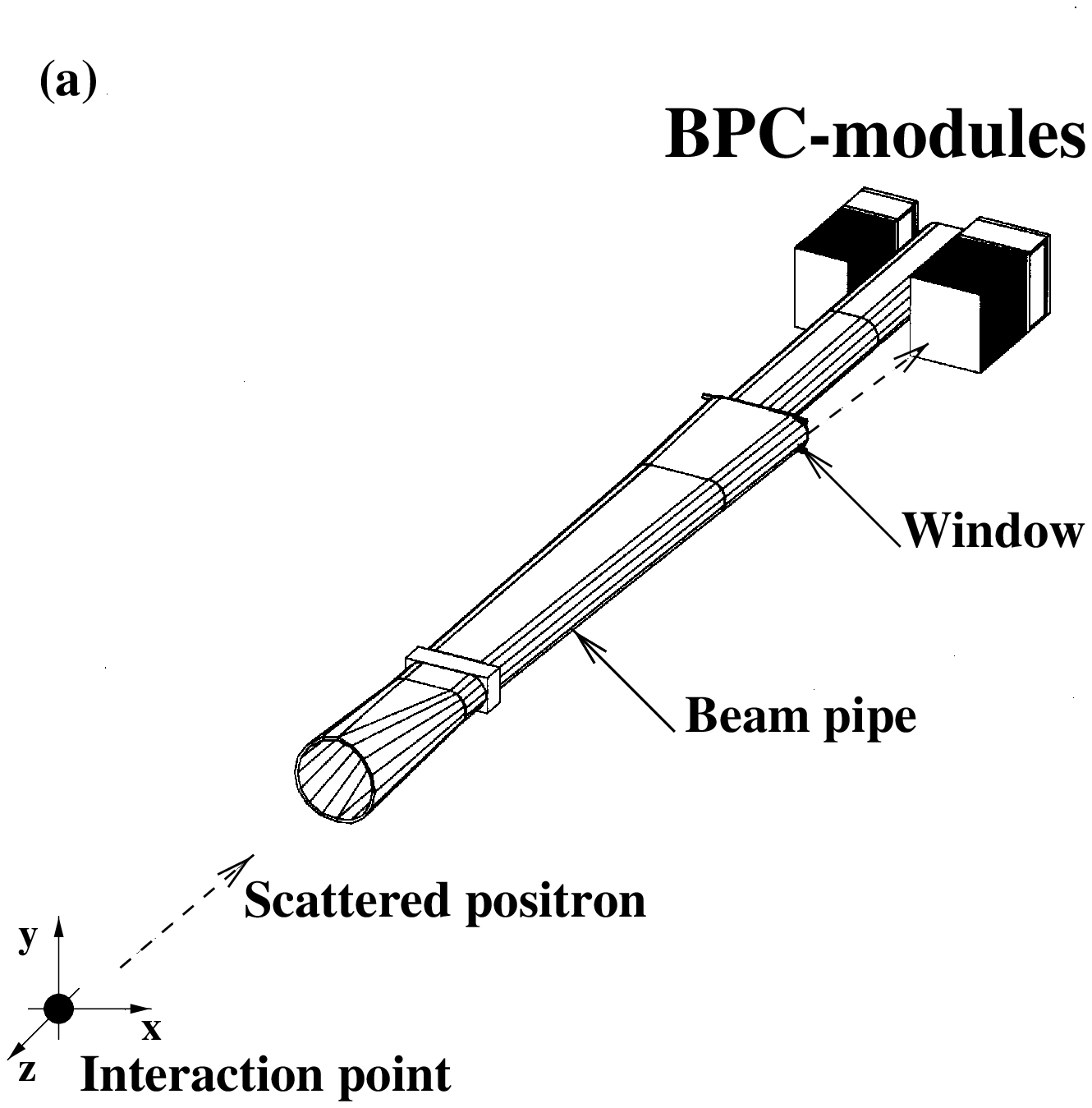,width=0.46\textwidth,height=0.46\textwidth}}}
\put (0.5,0.5){\mbox{\epsfig{file=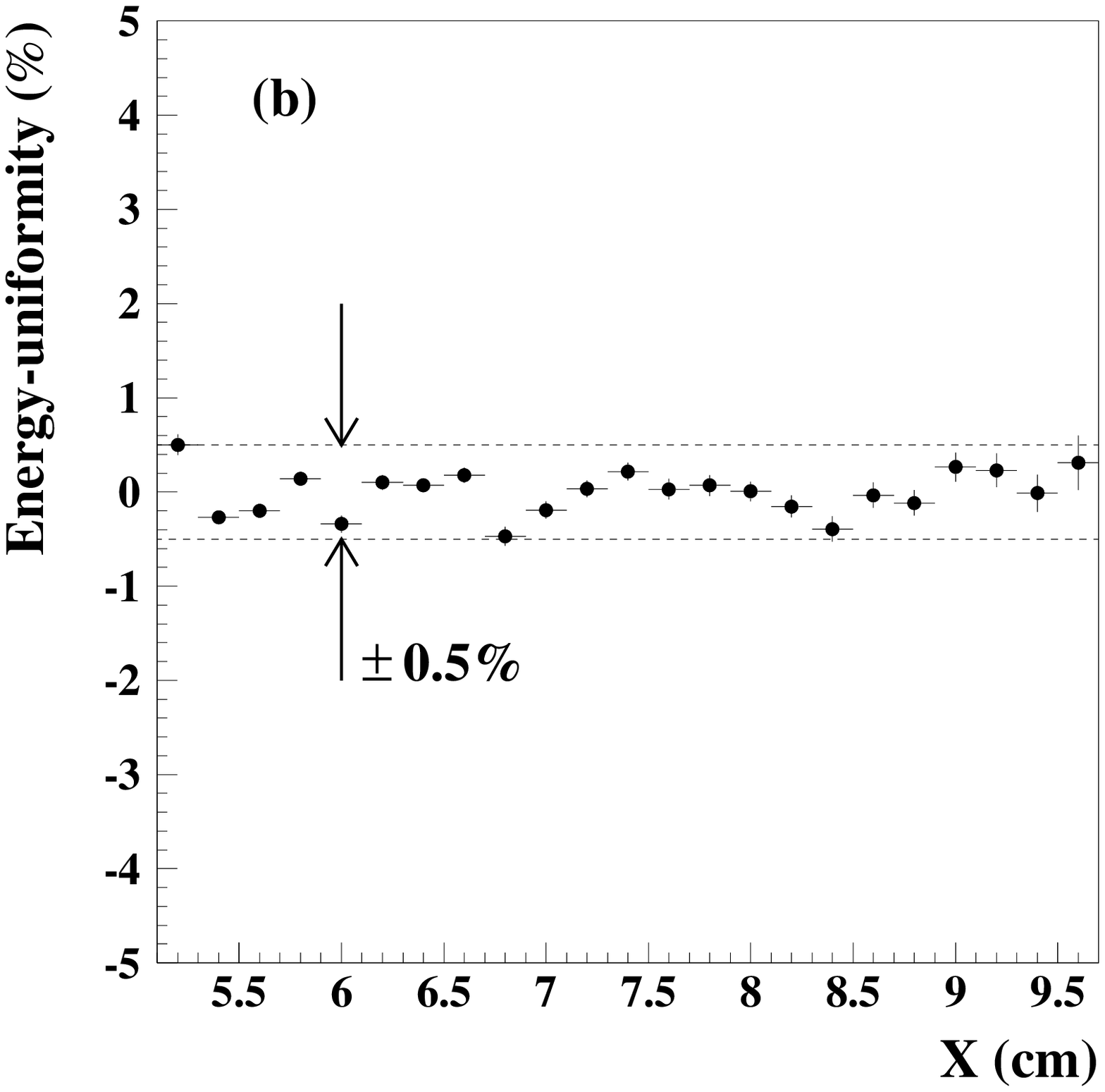,width=0.46\textwidth,height=0.46\textwidth}}}
\put (0,0){\mbox{\epsfig{file=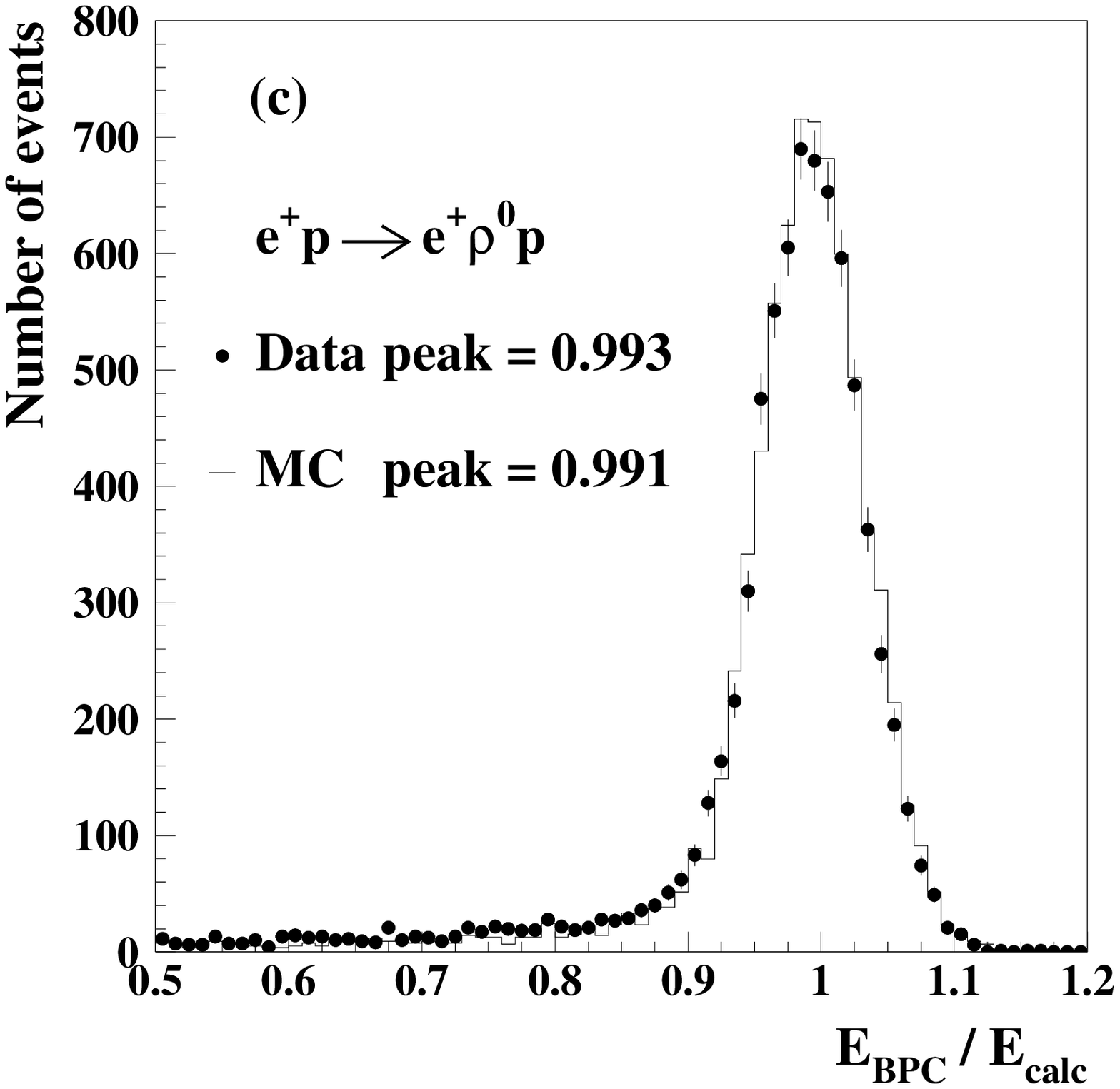,width=0.461\textwidth,height=0.461\textwidth}}}
\put (0.5,0){\mbox{\epsfig{file=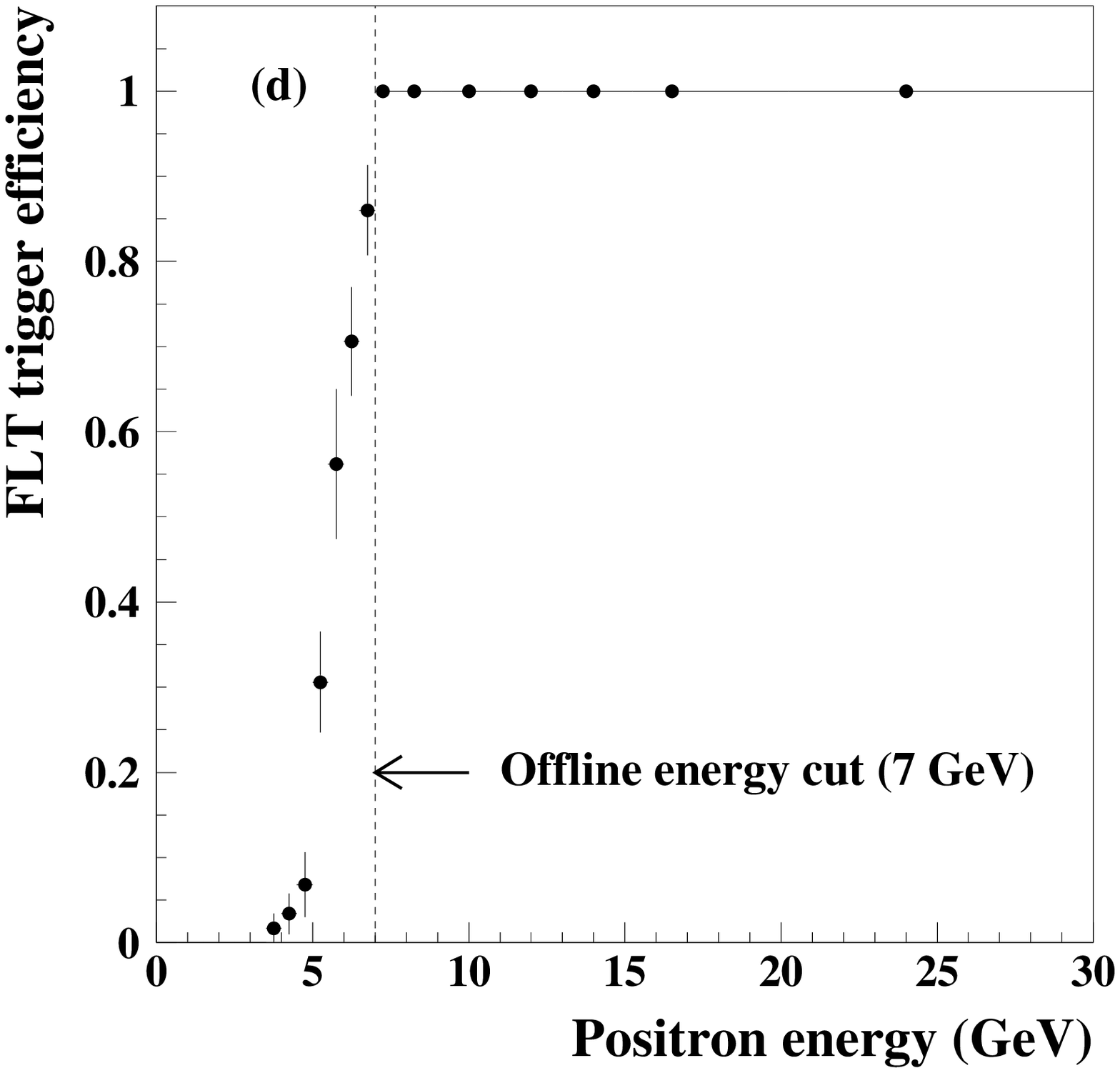,width=0.46\textwidth,height=0.46\textwidth}}}
\end{picture}
\caption{ (a) Schematic layout of the BPC and the beampipe.  The 
BPC is located at Z=-2937 mm, and the inner edge is at X=44 mm. 
(b) The fractional deviation from the mean energy for KP events 
as a function of the scattered positron X impact position at the 
BPC. (c) A comparison between MC and data of the ratio of the 
measured positron energy in the BPC, E$_{BPC}$, to the calculated
positron energy, E$_{calc}$ for elastic $\rho^0$ events.  (d) The
FLT trigger efficiency as a function of BPC energy.} 
\label{fig:uniformity} \end{figure}

\begin{figure}[t]
\setlength{\unitlength}{\textwidth}
\begin{picture} (1,1)
\put (0.35,0.975){\mbox{\Huge\bf ZEUS 1995}}
\put (0,0.5){\mbox{\epsfig{file=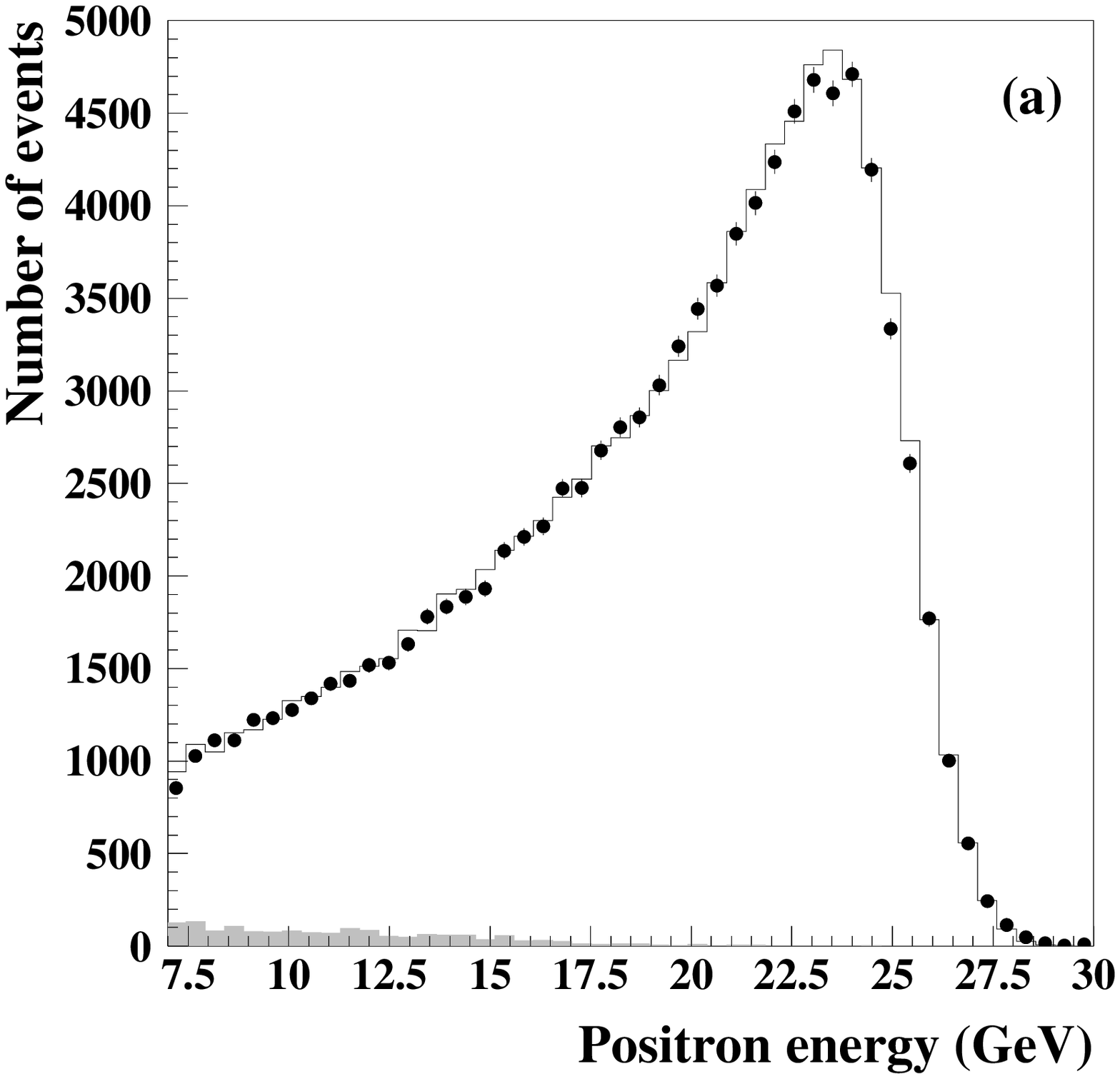,width=0.46\textwidth,height=0.46\textwidth}}}
\put (0.5,0.5){\mbox{\epsfig{file=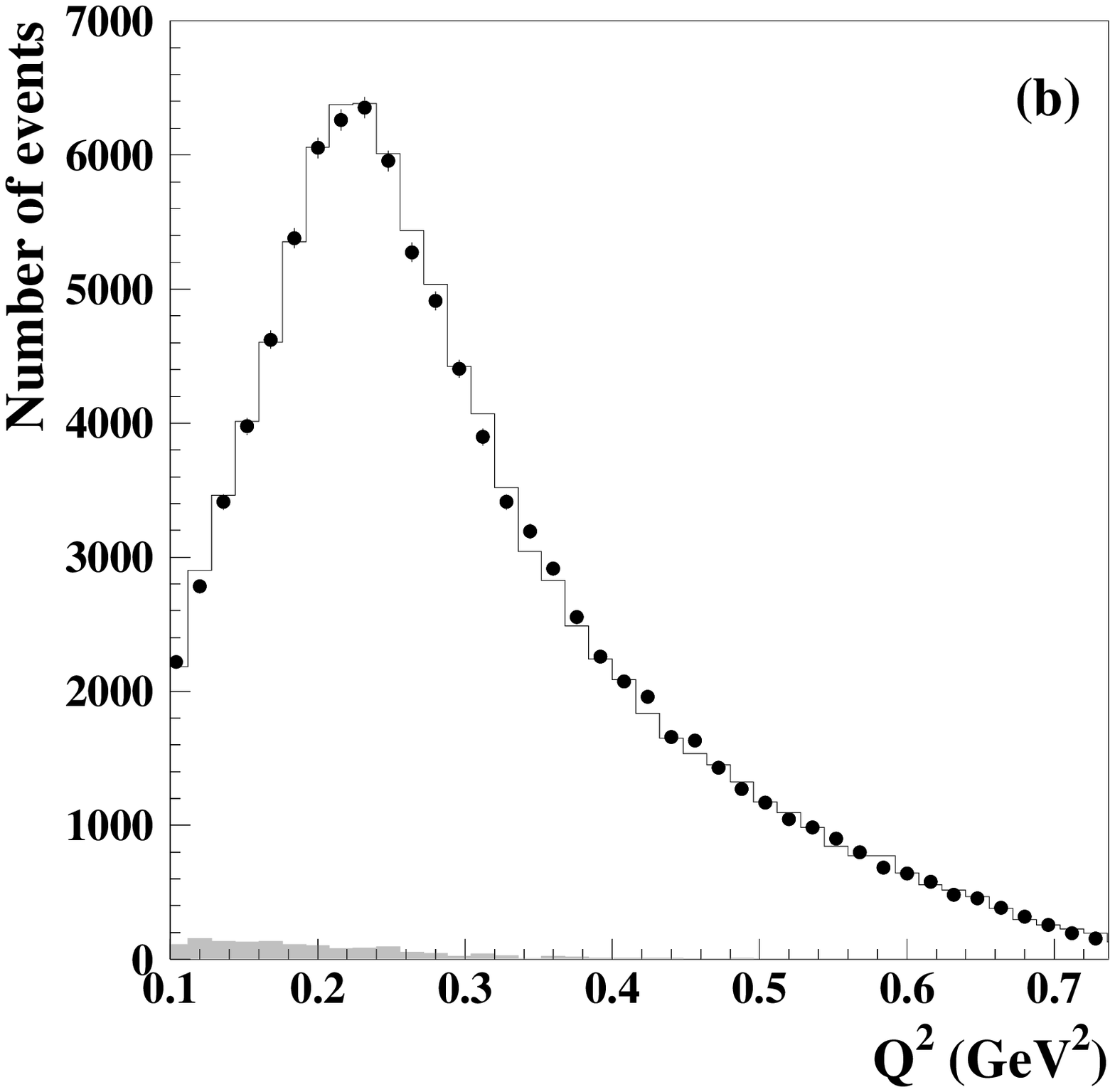,width=0.46\textwidth,height=0.46\textwidth}}}
\put (0,0){\mbox{\epsfig{file=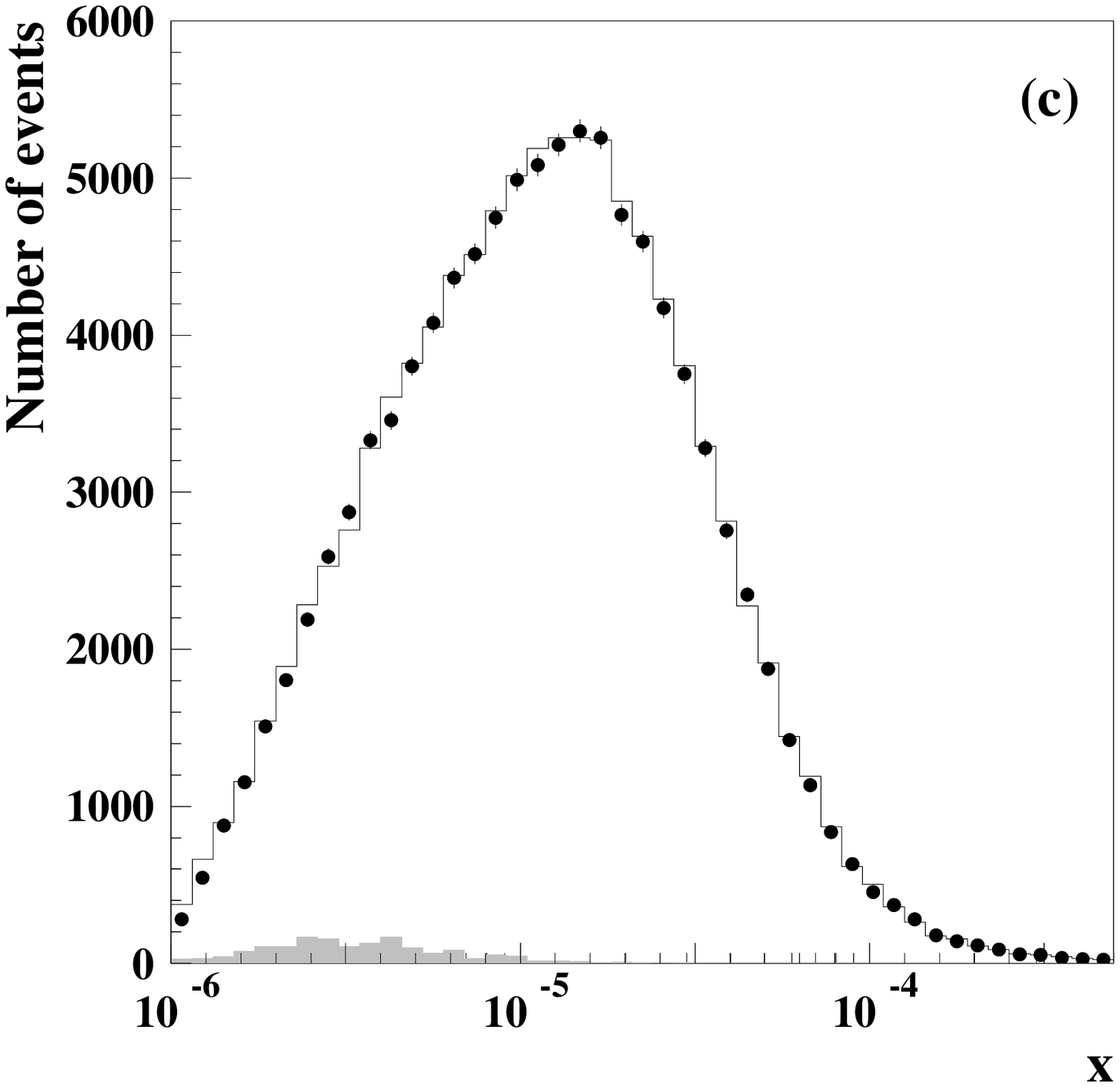,width=0.46\textwidth,height=0.46\textwidth}}}
\put (0.5,0){\mbox{\epsfig{file=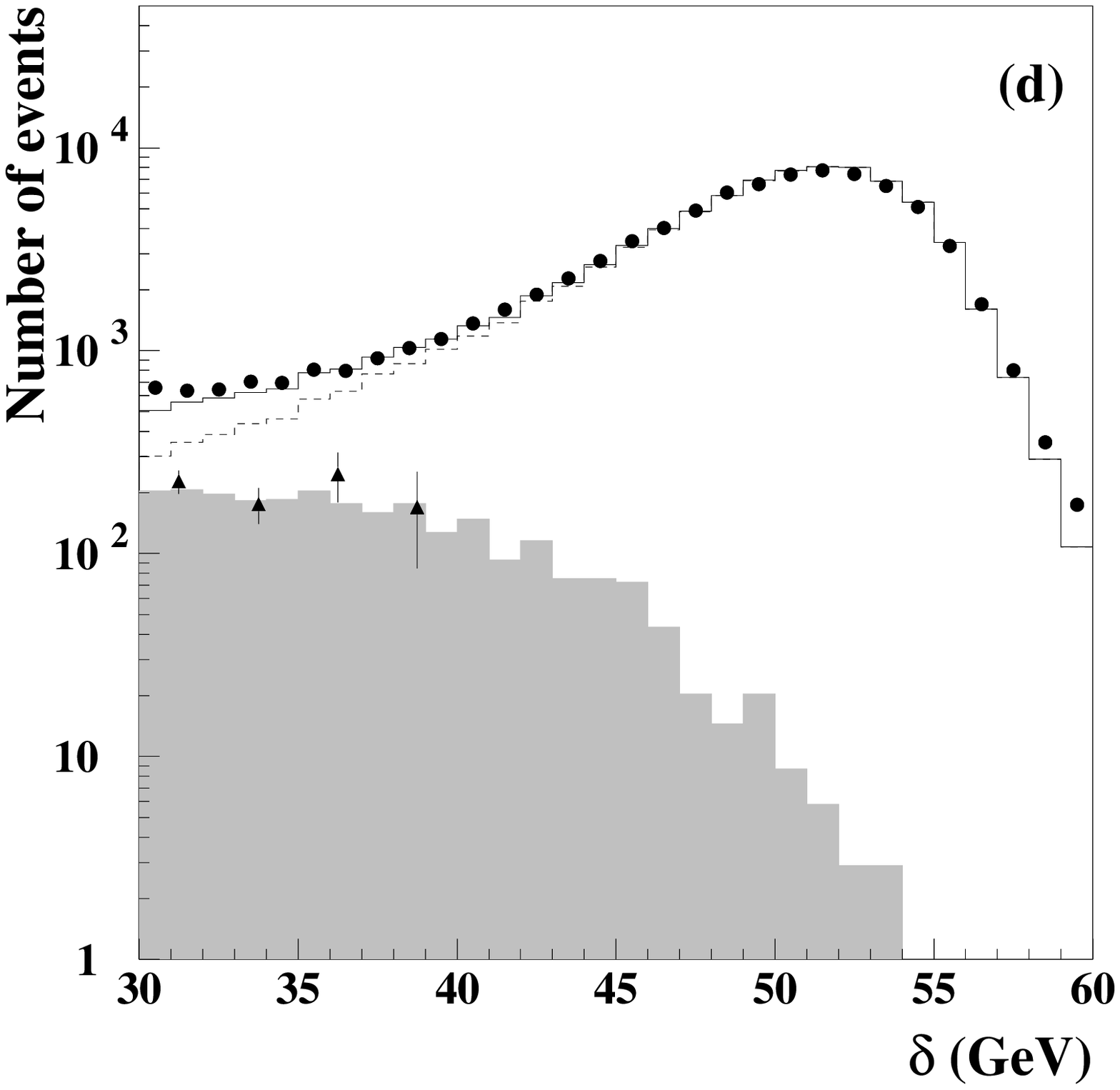,width=0.46\textwidth,height=0.46\textwidth}}}
\end{picture}
\caption{Comparisons between simulated and data distributions for the 
variables:  (a) scattered positron energy;  (b) $Q^2$;  (c) $x$; and  (d) 
$\delta$ . Data are shown as solid circles,
photoproduction simulation as shaded regions, and the sum of the 
signal and photoproduction simulations as solid lines. In figure (d) 
the signal simulation is shown as a dashed line and 
the measured background as triangular points. 
The structure function in the simulation has been 
reweighted to that measured in the present analysis.}
\label{fig:empz}
\end{figure}

\begin{figure}[t]
\setlength{\unitlength}{\textwidth}
\begin{picture} (1.0,1.25) (0,0)
\put (0.35,1.25){\mbox{\Huge\bf ZEUS 1995}}
\mbox{\epsfig{file=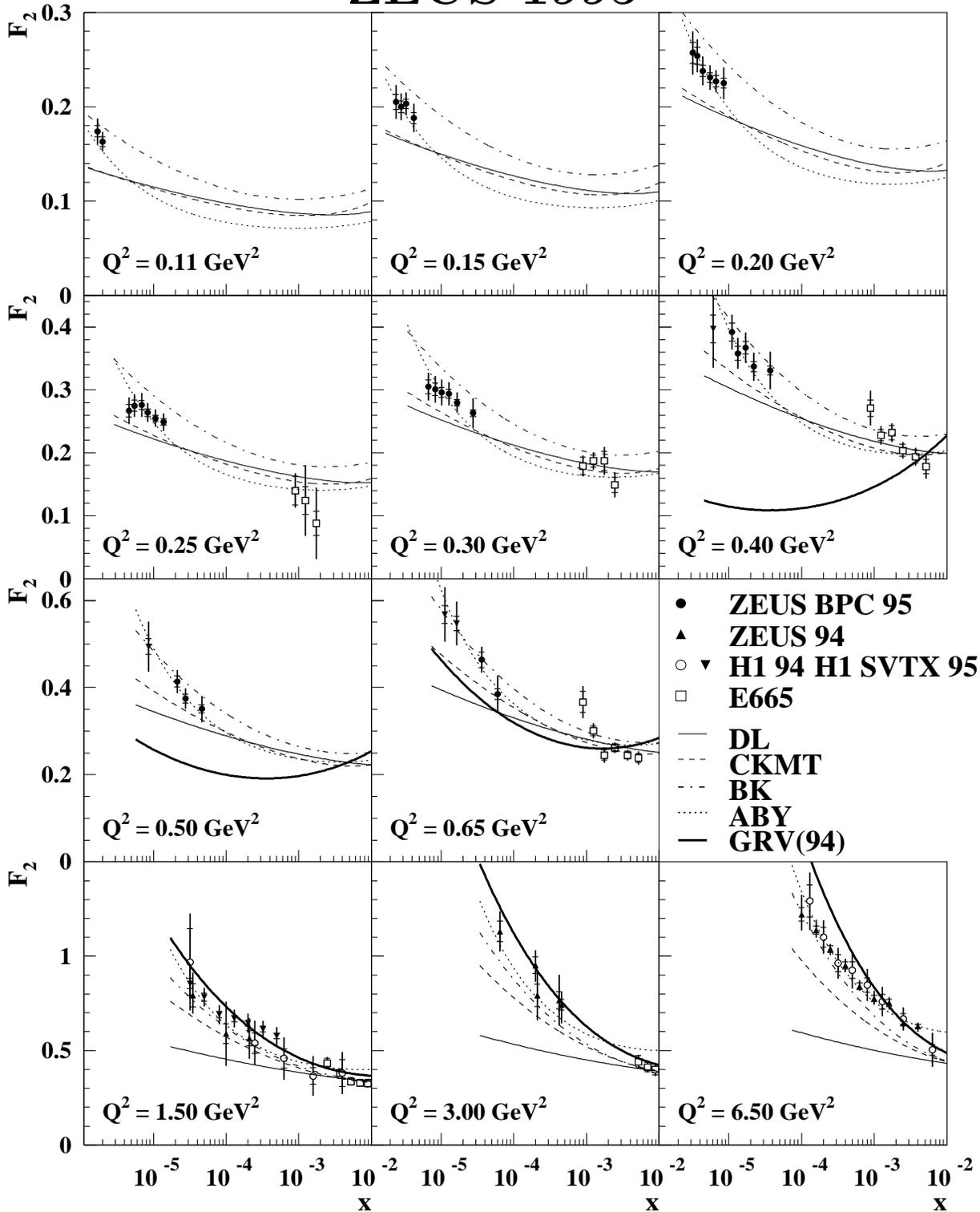,width=\textwidth,height=1.25\textwidth}}
\end{picture}
\caption{$F_2 (x, Q^2)$ as a function of $x$ for different Q$^2$ bins. 
The data from this analysis, ZEUS BPC 95, are shown as solid dots, with E665, 
H1 and previous ZEUS points shown as open squares, open circles 
and solid triangles, respectively. New points from 
H1\cite{h1-svtx} at low $Q^2$ are also shown as solid inverted 
triangles (the point at $Q^2=0.35$ GeV$^2$ is displayed in the 
0.4 GeV$^2$ panel; additional points at 2.5 and 3.5 GeV$^2$
not shown.) Curves from the models of 
DL, CKMT, BK, ABY and GRV are overlaid.} \label{fig:f2_x_theory} 
\end{figure}

\begin{figure}[ht]
\setlength{\unitlength}{\textwidth}
\begin{picture} (1.0,1.25) (0,0)
\put (0.35,1.255){\mbox{\Huge\bf ZEUS 1995}}
\put (0.0,0.0){\mbox{\epsfig{file=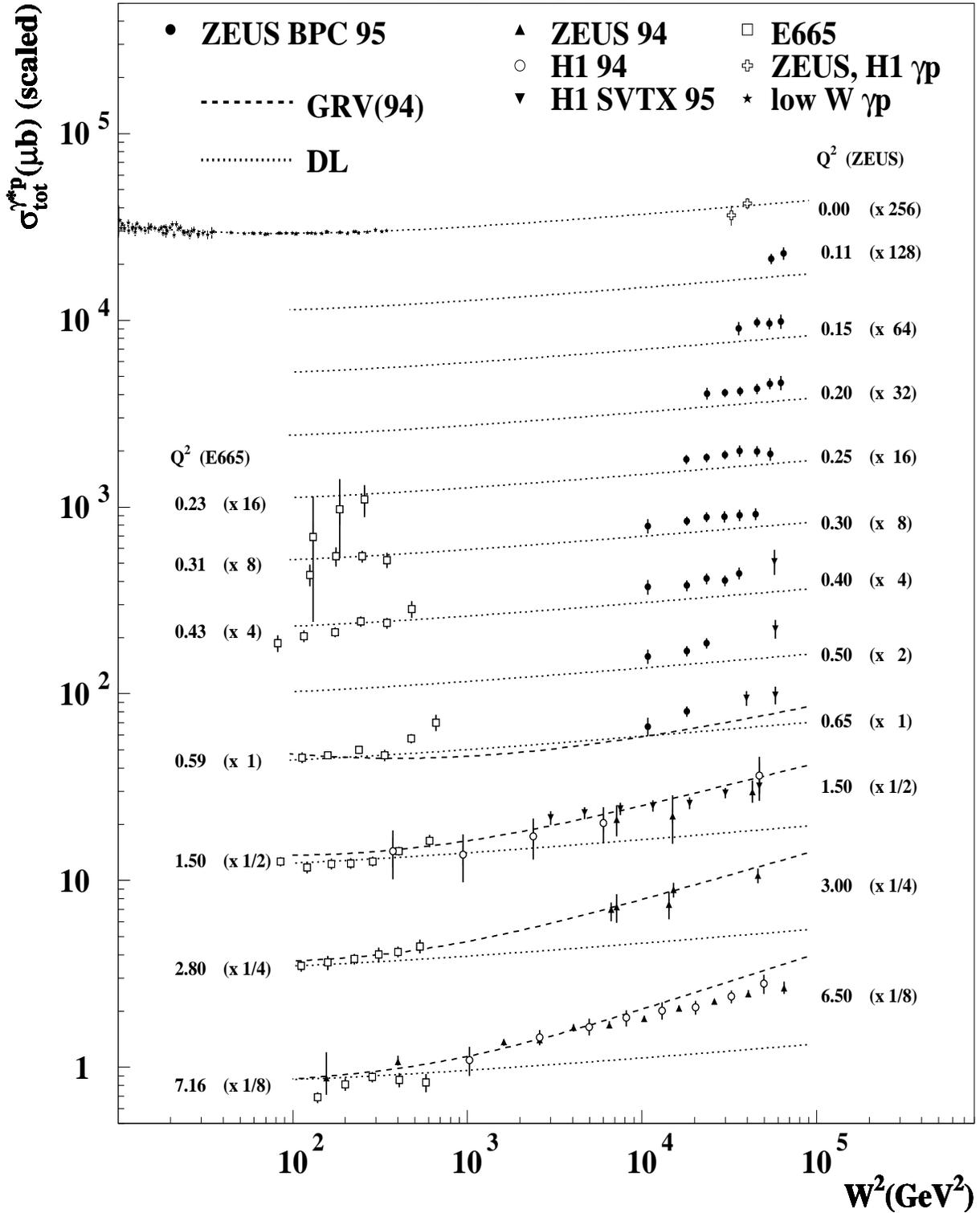,height=1.25\textwidth,width=\textwidth}}}
\end{picture}
\caption{The total virtual photon-proton cross-section 
$\sigma_{tot}^{\gamma^*p}$ as a function of $W^2$. The data from 
this analysis  (ZEUS BPC 95), previous 1994 ZEUS analyses, H1, 
and E665 are shown. The total cross-section for real photon-proton 
scattering from ZEUS, H1 and photoproduction experiments
at low $W$ are also shown. The predictions of DL 
and GRV (at the ZEUS $Q^2$ values) 
are indicated by the dotted and dashed curves respectively.}
\label{fig:sig_w2}
\end{figure}

\clearpage
\newpage

\appendix
\newcounter{subtab}
\setcounter{subtab}{1}
\newcounter{aptab}
\setcounter{aptab}{1}
\setcounter{table}{0}
\renewcommand{\thetable}{\Alph{aptab}}

\tabcolsep 3pt
\begin{table}[P]
\begin{sideways}\begin{minipage}[b]{\textheight}
{\Large \bf Appendix A: individual effects of systematic uncertainties}
\tiny
\begin{center}
\vspace{2.5cm} 
\hspace{-2.5cm}\begin{tabular}{|c|c|c|c|c|c|c|c|c|c|c|c|c|c|c|c|c|c|c|c|c|c|c|c|c|c|c|}
\multicolumn{1}{c}{ \begin{rotate}{45} $Q^2$ (GeV$^2$)\end{rotate}} 
& \multicolumn{1}{c}{ \begin{rotate}{45} $y$ \end{rotate}} 
& \multicolumn{1}{c}{ \begin{rotate}{45} nominal $F_2$ \end{rotate}} 
& \multicolumn{1}{c}{ \begin{rotate}{45}BPC energy scale + \end{rotate}} 
& \multicolumn{1}{c}{ \begin{rotate}{45}BPC energy scale - \end{rotate}} 
& \multicolumn{1}{c}{ \begin{rotate}{45}CAL energy scale +  \end{rotate}}   
& \multicolumn{1}{c}{ \begin{rotate}{45}CAL energy scale -  \end{rotate}}   
& \multicolumn{1}{c}{ \begin{rotate}{45}$y_{JB}$ cut + \end{rotate}}   
& \multicolumn{1}{c}{ \begin{rotate}{45}$y_{JB}$ cut - \end{rotate}}   
& \multicolumn{1}{c}{ \begin{rotate}{45}positron identification +\end{rotate}}   
& \multicolumn{1}{c}{ \begin{rotate}{45}positron identification -\end{rotate}}   
& \multicolumn{1}{c}{ \begin{rotate}{45}non-linearity of BPC energy scale +  \end{rotate}}   
& \multicolumn{1}{c}{ \begin{rotate}{45}non-linearity of BPC energy scale -  \end{rotate}}   
& \multicolumn{1}{c}{ \begin{rotate}{45} BPC position alignment + \end{rotate}}   
& \multicolumn{1}{c}{ \begin{rotate}{45} BPC position alignment - \end{rotate}}   
& \multicolumn{1}{c}{ \begin{rotate}{45} CAL noise cuts +\end{rotate}}   
& \multicolumn{1}{c}{ \begin{rotate}{45} CAL noise cuts -\end{rotate}}   
& \multicolumn{1}{c}{ \begin{rotate}{45}  $\delta$ cut +\end{rotate}}   
& \multicolumn{1}{c}{ \begin{rotate}{45}  $\delta$ cut -\end{rotate}}   
& \multicolumn{1}{c}{ \begin{rotate}{45}photoproduction background +\end{rotate}}   
& \multicolumn{1}{c}{ \begin{rotate}{45}photoproduction background -\end{rotate}}   
& \multicolumn{1}{c}{ \begin{rotate}{45}fraction of diffractive events + \end{rotate}}   
& \multicolumn{1}{c}{ \begin{rotate}{45}fraction of diffractive events - \end{rotate}}   
& \multicolumn{1}{c}{ \begin{rotate}{45}acceptance from the simulation +\end{rotate}}   
& \multicolumn{1}{c}{ \begin{rotate}{45}acceptance from the simulation -\end{rotate}}   
& \multicolumn{1}{c}{ \begin{rotate}{45}radiative corrections +\end{rotate}}   
& \multicolumn{1}{c}{ \begin{rotate}{45}radiative corrections -\end{rotate}}   
\\ \hline
\rule{0mm}{2.5mm}0.11 & 0.60 & 0.163 &   0.91 &  -0.91 &  -1.03 &   1.03 &  -0.03 &   0.03 &   0.20 &  -0.20 &  -2.26 &   2.26 &
\rule{0mm}{2.5mm}-1.55 &   1.55 &  -0.01 &   0.01 &   1.18 &  -1.18 &   1.34 &  -1.34 &   1.10 &  -1.10 &   4.00 &  -4.00 &   3.82 &   3.82 \\  
\rule{0mm}{2.5mm}0.11 & 0.70 & 0.174 &   1.20 &  -1.20 &  -2.48 &   2.48 &   0.03 &  -0.03 &   0.41 &  -0.41 &  -1.82 &   1.82 &
\rule{0mm}{2.5mm}2.22 &  -2.22 &   0.05 &  -0.05 &  -1.62 &   1.62 &   2.51 &  -2.51 &   1.41 &  -1.41 &   5.00 &  -5.00 &   3.40 &   3.40 \\  
\rule{0mm}{2.5mm}0.15 & 0.40 & 0.188 &   0.24 &  -0.24 &  -0.42 &   0.42 &  -0.14 &   0.14 &   0.65 &  -0.65 &  -1.53 &   1.53 &
6.15 &   6.15 &  -0.13 &   0.13 &   0.03 &  -0.03 &   0.60 &  -0.60 &   0.28 &  -0.28 &   2.00 &  -2.00 &   4.00 &   4.00 \\  
\rule{0mm}{2.5mm}0.15 & 0.50 & 0.203 &   1.53 &  -1.53 &  -0.46 &   0.46 &  -0.22 &   0.22 &   1.18 &  -1.18 &  -2.98 &   2.98 &
0.52 &  -0.52 &   0.44 &  -0.44 &   1.93 &  -1.93 &   0.78 &  -0.78 &   0.21 &  -0.21 &   3.00 &  -3.00 &   4.03 &   4.03 \\  
\rule{0mm}{2.5mm}0.15 & 0.60 & 0.200 &  -0.66 &   0.66 &  -0.95 &   0.95 &  -0.18 &   0.18 &   0.78 &  -0.78 &  -0.16 &   0.16 &
3.53 &  -3.53 &   0.03 &  -0.03 &   0.47 &  -0.47 &   2.01 &  -2.01 &   0.91 &  -0.91 &   4.00 &  -4.00 &   3.82 &   3.82 \\  
\rule{0mm}{2.5mm}0.15 & 0.70 & 0.205 &   1.17 &  -1.17 &  -1.80 &   1.80 &  -0.10 &   0.10 &  -0.27 &   0.27 &  -2.22 &   2.22 &
2.77 &  -2.77 &  -0.77 &   0.77 &  -1.42 &   1.42 &   3.53 &  -3.53 &   2.24 &  -2.24 &   5.00 &  -5.00 &   3.40 &   3.40 \\  
\rule{0mm}{2.5mm}0.20 & 0.26 & 0.225 &   2.62 &  -2.62 &  -0.43 &   0.43 &  -0.01 &   0.01 &  -0.18 &   0.18 &  -2.94 &   2.94 &
4.21 &   4.21 &   0.02 &  -0.02 &  -0.03 &   0.03 &   0.32 &  -0.32 &   0.42 &  -0.42 &   2.00 &  -2.00 &   3.60 &   3.60 \\  
\rule{0mm}{2.5mm}0.20 & 0.33 & 0.228 &   0.97 &  -0.97 &  -0.19 &   0.19 &  -0.46 &   0.46 &   0.30 &  -0.30 &  -0.86 &   0.86 &
0.76 &  -0.76 &  -0.32 &   0.32 &  -0.03 &   0.03 &   0.45 &  -0.45 &   0.19 &  -0.19 &   2.00 &  -2.00 &   3.86 &   3.86 \\  
\rule{0mm}{2.5mm}0.20 & 0.40 & 0.231 &   0.31 &  -0.31 &  -0.32 &   0.32 &  -0.38 &   0.38 &   0.55 &  -0.55 &  -2.21 &   2.21 &
2.29 &  -2.29 &  -0.37 &   0.37 &   0.22 &  -0.22 &   0.88 &  -0.88 &   0.33 &  -0.33 &   2.00 &  -2.00 &   4.00 &   4.00 \\  
\rule{0mm}{2.5mm}0.20 & 0.50 & 0.238 &   0.35 &  -0.35 &  -0.43 &   0.43 &  -0.37 &   0.37 &   0.58 &  -0.58 &  -2.30 &   2.30 &
2.67 &  -2.67 &  -0.41 &   0.41 &  -0.51 &   0.51 &   1.36 &  -1.36 &   1.02 &  -1.02 &   3.00 &  -3.00 &   4.03 &   4.03 \\  
\rule{0mm}{2.5mm}0.20 & 0.60 & 0.254 &   1.12 &  -1.12 &  -0.77 &   0.77 &  -0.21 &   0.21 &  -0.77 &   0.77 &  -2.52 &   2.52 &
1.89 &  -1.89 &  -0.44 &   0.44 &  -0.16 &   0.16 &   1.67 &  -1.67 &   0.78 &  -0.78 &   4.00 &  -4.00 &   3.82 &   3.82 \\  
\rule{0mm}{2.5mm}0.20 & 0.70 & 0.257 &   1.39 &  -1.39 &  -2.21 &   2.21 &  -0.11 &   0.11 &  -0.29 &   0.29 &  -2.86 &   2.86 &
1.27 &  -1.27 &  -1.50 &   1.50 &  -0.62 &   0.62 &   3.14 &  -3.14 &   0.80 &  -0.80 &   5.00 &  -5.00 &   3.40 &   3.40 \\  
\rule{0mm}{2.5mm}0.25 & 0.20 & 0.249 &   1.75 &  -1.75 &  -0.92 &   0.92 &  -0.20 &   0.20 &  -0.64 &   0.64 &  -0.90 &   0.90 &
0.91 &  -0.91 &  -0.18 &   0.18 &   0.00 &   0.00 &   0.37 &  -0.37 &   0.95 &  -0.95 &   2.00 &  -2.00 &   3.30 &   3.30 \\  
\rule{0mm}{2.5mm}0.25 & 0.26 & 0.256 &   1.65 &  -1.65 &  -0.22 &   0.22 &  -0.33 &   0.33 &  -0.77 &   0.77 &  -1.86 &   1.86 &
1.83 &  -1.83 &  -0.28 &   0.28 &   0.00 &   0.00 &   0.43 &  -0.43 &   0.50 &  -0.50 &   2.00 &  -2.00 &   3.60 &   3.60 \\  
\rule{0mm}{2.5mm}0.25 & 0.33 & 0.264 &   0.70 &  -0.70 &  -0.29 &   0.29 &  -0.43 &   0.43 &  -0.90 &   0.90 &  -2.81 &   2.81 &
1.98 &  -1.98 &  -0.43 &   0.43 &  -0.01 &   0.01 &   0.67 &  -0.67 &   0.61 &  -0.61 &   2.00 &  -2.00 &   3.86 &   3.86 \\  
\rule{0mm}{2.5mm}0.25 & 0.40 & 0.276 &   1.60 &  -1.60 &  -0.02 &   0.02 &  -0.79 &   0.79 &  -0.53 &   0.53 &  -3.55 &   3.55 &
2.71 &  -2.71 &  -0.45 &   0.45 &   0.15 &  -0.15 &   1.05 &  -1.05 &   0.45 &  -0.45 &   2.00 &  -2.00 &   4.00 &   4.00 \\  
\rule{0mm}{2.5mm}0.25 & 0.50 & 0.275 &  -0.02 &   0.02 &  -0.56 &   0.56 &  -0.37 &   0.37 &  -0.23 &   0.23 &  -1.90 &   1.90 &
3.46 &  -3.46 &  -0.33 &   0.33 &   0.40 &  -0.40 &   1.16 &  -1.16 &   0.67 &  -0.67 &   3.00 &  -3.00 &   4.03 &   4.03 \\  
\rule{0mm}{2.5mm}0.25 & 0.60 & 0.267 &  -0.60 &   0.60 &  -1.38 &   1.38 &  -0.26 &   0.26 &  -0.49 &   0.49 &  -0.15 &   0.15 &
3.47 &  -3.47 &   0.22 &  -0.22 &   1.05 &  -1.05 &   3.14 &  -3.14 &   1.16 &  -1.16 &   4.00 &  -4.00 &   3.82 &   3.82 \\  
\rule{0mm}{2.5mm}0.30 & 0.12 & 0.263 &  -1.42 &   1.42 &  -4.24 &   4.24 &   1.41 &  -1.41 &   0.57 &  -0.57 &   4.14 &  -4.14 &
1.57 &  -1.57 &   2.62 &  -2.62 &  -0.04 &   0.04 &  -0.26 &   0.26 &   3.20 &  -3.20 &   2.00 &  -2.00 &   2.76 &   2.76 \\  
\rule{0mm}{2.5mm}0.30 & 0.20 & 0.280 &   0.56 &  -0.56 &  -0.74 &   0.74 &  -0.12 &   0.12 &  -1.01 &   1.01 &   0.21 &  -0.21 &
2.75 &  -2.75 &  -0.25 &   0.25 &   0.04 &  -0.04 &   0.39 &  -0.39 &   1.35 &  -1.35 &   2.00 &  -2.00 &   3.30 &   3.30 \\  
\rule{0mm}{2.5mm}0.30 & 0.26 & 0.295 &   1.42 &  -1.42 &  -0.17 &   0.17 &  -0.41 &   0.41 &  -1.12 &   1.12 &  -2.04 &   2.04 &
2.65 &  -2.65 &  -0.40 &   0.40 &   0.05 &  -0.05 &   0.53 &  -0.53 &   0.26 &  -0.26 &   2.00 &  -2.00 &   3.60 &   3.60 \\  
\rule{0mm}{2.5mm}0.30 & 0.33 & 0.296 &   1.53 &  -1.53 &   0.00 &   0.00 &  -0.79 &   0.79 &  -0.69 &   0.69 &  -3.22 &   3.22 &
2.73 &  -2.73 &  -0.70 &   0.70 &   0.10 &  -0.10 &   0.67 &  -0.67 &  -0.19 &   0.19 &   2.00 &  -2.00 &   3.86 &   3.86 \\  
\rule{0mm}{2.5mm}0.30 & 0.40 & 0.301 &   1.30 &  -1.30 &  -0.18 &   0.18 &  -0.53 &   0.53 &  -0.61 &   0.61 &  -2.41 &   2.41 &
2.88 &  -2.88 &  -0.55 &   0.55 &   0.07 &  -0.07 &   0.84 &  -0.84 &   0.48 &  -0.48 &   2.00 &  -2.00 &   4.00 &   4.00 \\  
\rule{0mm}{2.5mm}0.30 & 0.50 & 0.305 &   1.68 &  -1.68 &  -0.42 &   0.42 &  -0.35 &   0.35 &   1.20 &  -1.20 &  -3.44 &   3.44 &
1.49 &  -1.49 &  -0.04 &   0.04 &   0.97 &  -0.97 &   1.34 &  -1.34 &   0.46 &  -0.46 &   3.00 &  -3.00 &   4.03 &   4.03 \\  
\rule{0mm}{2.5mm}0.40 & 0.12 & 0.332 &  -1.54 &   1.54 &  -4.11 &   4.11 &   0.96 &  -0.96 &   0.36 &  -0.36 &   4.28 &  -4.28 &
2.42 &  -2.42 &   1.95 &  -1.95 &  -0.03 &   0.03 &  -0.22 &   0.22 &   2.94 &  -2.94 &   2.00 &  -2.00 &   2.76 &   2.76 \\  
\rule{0mm}{2.5mm}0.40 & 0.20 & 0.337 &   2.09 &  -2.09 &  -0.70 &   0.70 &  -1.19 &   1.19 &  -1.02 &   1.02 &  -1.75 &   1.75 &
2.77 &  -2.77 &  -0.85 &   0.85 &   0.10 &  -0.10 &   0.47 &  -0.47 &   0.76 &  -0.76 &   2.00 &  -2.00 &   3.30 &   3.30 \\  
\rule{0mm}{2.5mm}0.40 & 0.26 & 0.367 &   1.07 &  -1.07 &   0.01 &  -0.01 &  -0.20 &   0.20 &  -1.26 &   1.26 &  -1.37 &   1.37 &
2.77 &  -2.77 &  -0.24 &   0.24 &   0.12 &  -0.12 &   0.49 &  -0.49 &   0.10 &  -0.10 &   2.00 &  -2.00 &   3.60 &   3.60 \\  
\rule{0mm}{2.5mm}0.40 & 0.33 & 0.358 &   0.48 &  -0.48 &  -0.19 &   0.19 &  -0.81 &   0.81 &  -1.03 &   1.03 &  -2.14 &   2.14 &
2.38 &  -2.38 &  -0.71 &   0.71 &   0.12 &  -0.12 &   0.74 &  -0.74 &  -0.27 &   0.27 &   2.00 &  -2.00 &   3.86 &   3.86 \\  
\rule{0mm}{2.5mm}0.40 & 0.40 & 0.392 &   1.38 &  -1.38 &  -0.09 &   0.09 &  -0.49 &   0.49 &  -1.90 &   1.90 &  -2.69 &   2.69 &
1.58 &  -1.58 &  -0.47 &   0.47 &   0.10 &  -0.10 &   1.08 &  -1.08 &   0.01 &  -0.01 &   2.00 &  -2.00 &   4.00 &   4.00 \\  
\rule{0mm}{2.5mm}0.50 & 0.12 & 0.351 &  -1.35 &   1.35 &  -3.87 &   3.87 &  -0.16 &   0.16 &   0.28 &  -0.28 &   3.80 &  -3.80 &
2.78 &  -2.78 &   1.09 &  -1.09 &  -0.05 &   0.05 &  -0.24 &   0.24 &   2.64 &  -2.64 &   2.00 &  -2.00 &   2.76 &   2.76 \\  
\rule{0mm}{2.5mm}0.50 & 0.20 & 0.375 &   0.33 &  -0.33 &  -0.82 &   0.82 &  -0.79 &   0.79 &  -0.85 &   0.85 &   0.41 &  -0.41 &
2.69 &  -2.69 &  -0.53 &   0.53 &   0.11 &  -0.11 &   0.37 &  -0.37 &   0.77 &  -0.77 &   2.00 &  -2.00 &   3.30 &   3.30 \\  
\rule{0mm}{2.5mm}0.50 & 0.26 & 0.414 &   0.74 &  -0.74 &   0.16 &  -0.16 &  -0.84 &   0.84 &  -1.74 &   1.74 &  -0.34 &   0.34 &
2.09 &  -2.09 &  -0.37 &   0.37 &   0.14 &  -0.14 &   0.55 &  -0.55 &   0.59 &  -0.59 &   2.00 &  -2.00 &   3.60 &   3.60 \\  
\rule{0mm}{2.5mm}0.65 & 0.12 & 0.386 &  -2.83 &   2.83 &  -4.34 &   4.34 &   3.14 &  -3.14 &   0.44 &  -0.44 &   5.25 &  -5.25 &
2.61 &  -2.61 &   3.46 &  -3.46 &  -0.07 &   0.07 &  -0.18 &   0.18 &   2.60 &  -2.60 &   2.00 &  -2.00 &   2.76 &   2.76 \\  
\rule{0mm}{2.5mm}0.65 & 0.20 & 0.464 &   1.68 &  -1.68 &  -0.42 &   0.42 &   0.75 &  -0.75 &  -1.00 &   1.00 &  -1.04 &   1.04 &
1.35 &  -1.35 &   0.45 &  -0.45 &   0.22 &  -0.22 &   0.52 &  -0.52 &   0.67 &  -0.67 &   2.00 &  -2.00 &   3.30 &   3.30 \\  \hline
\end{tabular}
\caption{
The determination of the systematic errors was discussed in
section \ref{sec-systematics}.  $Q^2$, $y$, and measured $F_2$ values are 
listed in columns 1 to 3. The individual effects of each of the
systematic checks on $F_2$ are displayed (in percentages of $F_2$) in 
columns  4 to 27.
}
\label{tab-systematics}
\end{center}
\end{minipage}\end{sideways}
\end{table}
\end{document}